\documentclass[useAMS,usenatbib,graphicx]{mn2e}
\usepackage{times}
\usepackage{epsfig}

\title[GRB~051008: A long, spectrally-hard dust-obscured GRB]{GRB~051008: A long, spectrally-hard dust-obscured GRB in a Lyman-Break Galaxy at $z \approx 2.8^\ast$}

\author[A. A. Volnova et. al.]{
A. A.~Volnova,$^{1}$\thanks{email:~alinusss@gmail.com}
 A. S.~Pozanenko,$^1$\thanks{email:~apozanen@iki.rssi.ru} J. Gorosabel,$^2$ D. A. Perley,$^{3,4}$ D. D. Frederiks,$^5$ 
 \newauthor D. A. Kann,$^{6,7}$ V. V. Rumyantsev,$^8$ V. V. Biryukov,$^{8,9}$ O. Burkhonov,$^{10}$ A. J. Castro-Tirado,$^2$ 
 \newauthor P. Ferrero,$^{11,12}$ S. V. Golenetskii,$^5$ S. Klose,$^6$ V. M. Loznikov,$^1$ P. Yu. Minaev,$^1$ 
 \newauthor  B. Stecklum,$^6$ D. S. Svinkin,$^{5}$ A. E. Tsvetkova,$^5$ A. de Ugarte Postigo,$^{2,13}$ and M. V. Ulanov$^{5}$\\
$^{1}$Space Research Institute, 84/32 Profsoyuznaya Street, Moscow 117997, Russia\\
$^2$Instituto de Astrof\'{i}sica de Andaluc\'{i}a del Consejo Superior de Investigaciones Cientificas (IAA-CSIC), Camino Bajo de Huetor~50, E-18080 Granada, Spain\\
$^3$Department of Astronomy, California Institute of Technology, MC 249-17, 1200 East California Blvd., Pasadena, CA 91125, USA\\
$^4$Hubble Fellow\\
$^5$Ioffe Physical-Technical Institute, Politekhnicheskaya 26, St.~Petersburg 194021, Russia\\
$^6$Th\"{u}ringer Landessternwarte Tautenburg,Sternwarte 5, 07778 Tautenburg, Germany\\
$^7$Max-Planck-Institut f\"{u}r extraterrestrische Physik, Giessenbachstra\ss e 1, 85748 Garching, Germany\\
$^8$Crimean Astrophysical Observatory, Taras Shevchenko National University of Kyiv, Nauchny, Crimea 98409, Ukraine\\
$^9$Crimean Laboratory of the Sternberg Astronomical Institute, Nauchny, Crimea 98409, Ukraine\\
$^{10}$Institute of Astronomy, Academy of Sciences of Uzbekistan, Tashkent, Uzbekistan\\
$^{11}$Instituto de Astrofisica de Canarias (IAC), 38200 La Laguna, Tenerife, Spain \\
$^{12}$Departamento de Astrofisica, Universidad de La Laguna (ULL), 38205 La Laguna, Tenerife, Spain \\
$^{13}$Dark Cosmology Centre, Niels Bohr Institute, University of Copenhagen, Juliane Maries Vej 30, 2100 Copenhagen, Denmark \\
$\ast$Based on observations made with the Nordic Optical Telescope, operated on the island of La Palma jointly by Denmark, Finland, Iceland, Norway, \\
and Sweden, in the Spanish Observatorio del Roque de los Muchachos of the Instituto de Astrofisica de Canarias.
}
\begin{document}

\date{Accepted XXXX. Received XXXX.}
\pagerange{\pageref{firstpage}--\pageref{lastpage}} \pubyear{XXXX}
\label{firstpage}
\maketitle

\begin{abstract} We present observations
of the dark Gamma-Ray Burst GRB~051008 provided by \textit{Swift}/BAT,
\textit{Swift}/XRT, Konus-\textit{WIND}, \textit{INTEGRAL}/SPI-ACS
in the high-energy domain and the Shajn, \textit{Swift}/UVOT, Tautenburg,
NOT, Gemini and Keck~I telescopes in the optical and near-infrared bands.
The burst was detected only in gamma- and X-rays
and neither a prompt optical nor a radio afterglow were detected down to deep limits.
We identified the host galaxy of the burst, which is a typical Lyman-break Galaxy (LBG)
with $R$-magnitude of $24\fm06 \pm 0\fm10$. A redshift of the galaxy of
$z = 2.77_{-0.20}^{+0.15}$ is measured photometrically due to the presence
of a clear, strong Lyman-break feature.
The host galaxy is a small starburst galaxy with moderate intrinsic extinction
($A_V = 0.3$) and has a SFR of $\sim 60 M_{\sun}/$ yr typical for LBGs.
It is one of the few cases where a GRB
host has been found to be a classical Lyman-break galaxy.
Using the redshift we estimate the isotropic-equivalent radiated energy of the burst to
be $E_{iso} = (1.15 \pm 0.20) \times 10^{54}$~erg.
We also provide evidence in favour of the hypothesis that the darkness of GRB~051008
is due to local absorption resulting from a dense circumburst medium.
\end{abstract}


\section{Introduction}
Cosmic gamma-ray bursts (GRBs) are among the most powerful events
in the Universe. A detection of a GRB is expected to be followed
by the detection of an afterglow in other spectral ranges: X-ray,
optical, and radio emissions. Presently, a large number
of ground-based robotic telescopes search for optical afterglows
rapidly after GRBs have occurred. Also, on-board the dedicated
\textit{Swift} space observatory \citep{swift}, the
UltraViolet/Optical Telescope \citep[UVOT,][]{uvot} is used for
rapid follow-up of UV and optical afterglows. However, it is now
clear that for a substantial fraction of events (from
20\%, \citealt{cenko,greiner}, to 35\%, \citealt{melandri}) we are currently
unable to detect any optical afterglow, even with rapid follow-up
within the first minutes after the GRB \citep[see also][]{fynbo2,lazatti}.

Faintness or the complete lack of an  optical afterglow (OA) may
be caused by different factors. Most prosaically, a failure to detect the OA
may be due to the low limiting magnitude of the observations, such
as when only small robotic telescopes observe the GRB position
early on (in contrast to larger rapid follow-up telescopes, see
\citealt{greiner}).

A possible factor for a GRB to be dark is the immediate
neighbourhood of its source. The ``extinction scenario''
\citep{Taylor} assumes that the emission of the OA is strongly
absorbed in the host galaxy. Moreover, the GRB source may be
surrounded by a dense circumburst medium~\citep{paczynski,gala}.

Another potential origin of dark bursts is a high redshift.
Emission with wavelengths
shorter than $912(1+z)$ \AA\, in the observer frame
is efficiently absorbed due to the Lyman-cutoff
when it passes through the intergalactic
medium~\citep{lamb}\footnote{At higher redshifts the Ly$\alpha$
forest becomes effectively
opaque causing the break to be at $1217(1+z)$ \AA}. For
$z \geq 4$, this Lyman drop-out falls into the $R_C$ band in the
optical in which most rapid searches for GRB afterglows are
carried out. Such GRBs can only be localized through rapid, deep
near-infrared (NIR) follow-up, e.g., as in the case of GRB~080913
\citep{greiner2,perez}.

To determine whether the burst is genuinely dark, due to some
physical factors rather than to an inefficient search, it is
useful to compare the observed optical/near-IR detections or upper
limits with the values expected from the brightness of the X-ray
afterglow. This approach is based on the standard fireball
model~\citep{fireball}, where the afterglow spectrum produced by
the synchrotron radiation follows the simple power law $F_{\nu}
\sim \nu^{-\beta}$. \citet{Jacob} proposed to define dark bursts
using an optical to X-ray spectral index $\beta_{OX} < 0.5$, since in
log space it is the shallowest expected slope of the synchrotron
spectrum in the standard afterglow model. Alternatively, \citet{vdhorst2}
uses the value of the X-ray spectral index $\beta_X$, assuming for the
definition of dark bursts the condition $\beta_{OX} - \beta_X < 0.5$.

Finally, the burst may be intrinsically faint and the low optical
luminosity of its OA may be due to a low density of the
interstellar medium into which the relativistic outburst of the
GRB propagates~\citep[][note that in this case a GRB may
not be dark according to the Jakobsson criterion, as it may also
have a very faint X-ray afterglow]{fireball}.

Long GRBs are known to be linked to the deaths of massive stars,
and hence to star-formation activity \citep{jimenez}. As a result,
one may expect some GRBs to be located in dusty
environments, with high extinction values along the lines of sight
towards the bursts, or even high bulk extinction in the host
galaxies. \citet{melandri2008} found that about 50\% of
\textit{Swift} dark bursts show evidence
of a mild extinction. Further studies by \citet{perley} showed
that the majority of dark bursts require $A_V^{host} \geq 1$ mag,
with a few cases reaching $\sim 2-6$ mag. There are many notable
examples of highly extinguished events: GRB~980828 \citep{djor};
GRB~051022 \citep{castro}; GRB~061222A
\citep{perley}; GRB~070306 \citep{jaun}; GRB~070521
\citep{perley}; GRB~080325 \citep{hashi}; GRB~080607
\citep{chen2010,perley2011}; GRB~090417B \citep{holl}; GRB~100621A
\citep{greiner3}; GRB~110709B and GRB~111215A \citep{zaud}.

The study of properties of host galaxies
is one of the tools for investigating the environment in which the GRBs
form \citep[e.g.,][]{savag}.
Recent research on large populations of
dark-GRB host galaxies has shown that generally, they do not
differ from the host galaxies of GRBs that suffer from little dust
extinction \citep{perley}, with the ``darkness'' being mostly due
to local extinction around the progenitor in galaxies of low to
medium redshifts. But one can note that some very dark GRBs trace a
population of extremely red host galaxies
\citep{hunt,rossi,svens}, whereas highly extinguished (but still
detected) GRB afterglows are preferentially linked with more
massive host galaxies \citep{kruhler}. The host galaxies
of dark GRBs in general have large SFRs \citep{perley2013}.

Redshifts of detected host galaxies vary in a wide range from $z = 0.0085$
\citep[GRB~980425][]{980425} to $z = 4.667$ \citep[GRB~100219A][]{thone3}
with a median value of $z_{host}$ about 1.4\footnote{http://www.grbhosts.org/}.
About 20\% of discovered GRB host galaxies lie at redshifts more than 2.5.
The search for and spectroscopic observations of galaxies with $z \geq 2.5$
is a non-trivial problem. In these cases techniques of photometric redshift
estimation are useful. E.g., the Lyman limit of hydrogen at
912~\AA~ in the rest frame of a galaxy helps to find ordinary starburst
galaxies at $z \sim 3$ and at higher redshifts to identify Lyman break galaxies (LBGs). 
But to date there were only a few associations between GRBs and LBGs 
\citep{971214,malesani}.
Also if GRBs trace star-formation one should expect that a typical GRB host 
galaxy at $z>2.5$ will be under-luminous and fall below $L^{\ast}$ of the LBG 
luminosity function \citep{021004,lbg}.
In the case of dark bursts the discovery of the host galaxy is often the only way
to determine the distances to these sources and to try to obtain an insight into
their nature \citep[e.g.,][]{jacob2012,kruhler2012}.

In this paper we determine and discuss properties of GRB~051008 and
its host galaxy. We present detailed parameters of the prompt
$\gamma$-ray emission observed by Konus-\textit{WIND} in the energy
range up to 14~MeV which is essential for determining the
physical parameters $E_{p}$ and $E_{iso}$.
We also report new observations of the host
galaxy which were not published earlier in preliminary studies
\citep{meastbul,meannap}. In particular these observations in $U$,
$g^{\prime}$, $R$, $I$, and $Z$ bands provided by the
Keck~I telescope, in $K^{\prime}$ band provided by the Gemini North
telescope, and ultra-violet observations provided by
\textit{Swift}/UVOT helped us to determine a secure photometric
redshift of the galaxy. In the paper we use the following
notations and conventions: time-decay, photon, and spectral
indices are indicated with $\alpha$, $\Gamma$, and $\beta$,
following the standard convention $t^{-\alpha}$,
$N^{-\Gamma}_{ph}$, and $\nu^{-\beta}$, respectively. We use the
following cosmological parameters $H_0 = 71$ km/s/Mpc, $\Omega_M =
0.27$, $\Omega_{\Lambda} = 0.73$ and cosmological calculations of
Ned Wright's Cosmology Calculator \citep{NED}. If not stated
otherwise, all errors represent one standard deviation, and upper
limits are three standard deviations for the parameter of
interest.
%
\section{The GAMMA-RAY BURST GRB~051008}
\subsection{Detection of the burst}

The gamma-ray burst GRB 051008 was detected by the BAT telescope
of the \textit{Swift}\, space observatory at 16:33:21 UT on
October 8, 2005~\citep{marsh}. The burst occurred during a
telemetry downlink transmission and that is why the corresponding
alert was not received by the BACODINE $\gamma$-ray alert
network~\footnote{{\scriptsize S.~Barthelmy, {\tt
http://gcn.gsfc.nasa.gov/}}} until 10 minutes after the event.
The spacecraft did
not slew to the burst immediately because of the Earth-limb observing
constraint. This condition also prevented \textit{Swift}/BAT from
recording all gamma radiation from the event (see Section 2.2).
This burst was also detected at 16:33:18 UT by the WAM
$\gamma$-ray monitor on-board the \textit{Suzaku} observatory and
had a duration of 48~s in the energy range from 100~keV to 2~Mev.
The burst spectrum in this energy range can be fitted fairly well
by a CPL model with $dN/dE \sim E^{\Gamma} e^{-E/E_{c}}$ with
photon index $\Gamma = -1.24 \pm 0.15$, and a cut-off energy of
$E_{c} = 1535^{+1419}_{-561}$ keV \citep{ohno}.

GRB~051008 was also recorded by Konus-\textit{WIND} (see Section
2.2), as well as the SPI-ACS detector on-board \textit{INTEGRAL}
observatory (see Section 2.3), and the Mars Observer
spacecraft~\footnote{{\scriptsize{\tt
http://www.ssl.berkeley.edu/ipn3/masterli.txt}}}.
%
\subsection{Konus-\textit{WIND} observations}

GRB~051008 triggered detector S2 of the Konus-\textit{WIND}
$\gamma$-ray spectrometer~\citep{Aptekar1995} at $T_0$=16:33:21
UT. As derived from the \textit{WIND} spacecraft ephemerides and
the \textit{Swift} localization of the GRB source, the
corresponding Earth-crossing time is $T_{\oplus}$ = $T_0$+2.8~s.

\begin{figure}
\centering
\includegraphics[width=84mm]{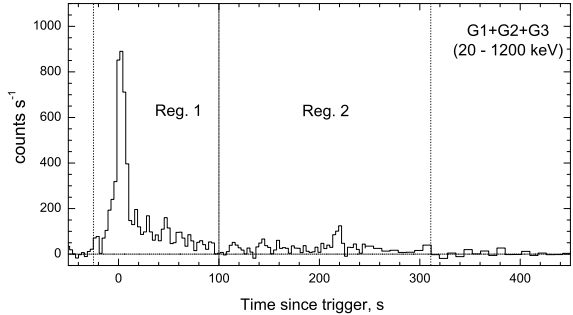}
\caption{An overview of the Konus-\textit{WIND} light curve. The
background-subtracted time history in the G1+G2+G3 bands is presented.
The fluence calculation intervals are tagged as \textit{Reg.~1}
and \textit{Reg.~2}. }
\label{FigLC}
\end{figure}
\begin{figure}
\centering
\includegraphics[height=84mm,angle=-90]{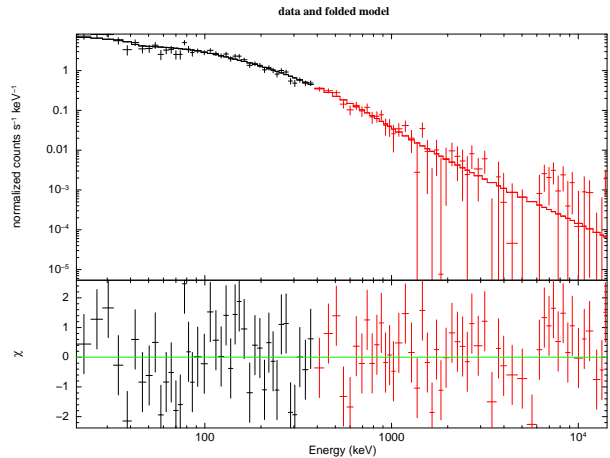}
\caption{The Konus-\textit{WIND} energy spectrum measured from
$T_0$ to $T_0$+8.2~s and its fit with the Band function
(see Table~\ref{TableSpec} for details).}
\label{FigSp15}
\end{figure}
\begin{figure}
\centering
\includegraphics[width=84mm]{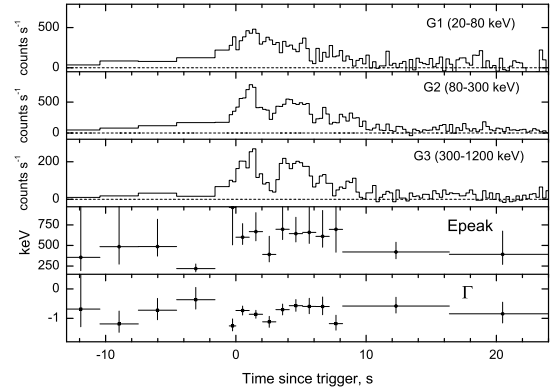}
\caption{Spectral evolution of the prompt $\gamma$-ray emission
during the main phase of the burst. Background-subtracted light
curves in three energy bands (G1, G2, G3) are shown, along with the
temporal behaviour of the CPL spectral model parameters $E_{peak}$
and $\Gamma$.}
\label{FigSpEvo}
\end{figure}

\begin{table}
\caption{GRB~051008 energy fluence measured with Konus-\textit{WIND}.}
\begin{tabular}{l|c|c}
\hline
Time interval & Spectral range, & Fluence, \\
(s) & keV & 10$^{-5}$ erg cm$^{-2}$ \\
\hline
Region~1$^a$ & 20 -- 14000 & 5.9 $\pm$ 1.0 \\
($T_0$-25 to $T_0$+98) & & \\
& & \\
Region~2 & 20 -- 14000 & 0.8 $\pm$ 0.5 \\
($T_0$+98 to $T_0$+311) & & \\
 & & \\
Whole burst & 20 -- 14000 & 6.7 $\pm$ 1.1 \\
($T_0$-25 to $T_0$+311) & & \\
\hline
\end{tabular}
\label{TableFluence}
\newline $^a${\footnotesize{The fluence for the pre-trigger part
(from $T_0$-25~s to $T_0$) is  estimated using a product of the
count fluence for this part and the conversion factor from a count
fluence to an energy fluence using the spectral information of the
time-averaged spectrum.}}
\newline $^b${\footnotesize{All the quoted errors are at the 90\%
confidence level.}}
\end{table}

\begin{table} %
\caption{Spectral lag between Konus-\textit{WIND} light curves in
the time interval [0 -- 10 s]}
\begin{tabular}{l|c|c}
\hline
Light curves & Time scale & $\tau_{lag}^a$ \\
 & (ms) & (s) \\
\hline
G3 -- G1& 256 & -0.2 $\pm$ 0.3$^b$ \\
 & & \\
G2 -- G1& 256 & -0.16 $\pm$ 0.14 \\
 & & \\
G3 -- G2& 64 & 0.06 $\pm$ 0.06 \\
\hline
\end{tabular}
\label{TableLag}

$^a${\footnotesize{In this table positive spectral lag means that
the spectrum has hard-to-soft evolution.}}
$^b${\footnotesize{All the quoted errors are at 1~$\sigma$ level.}}
\end{table}

\begin{table*}
 \begin{minipage}{140mm}
  \caption{Summary of Konus-\textit{WIND} spectral fits in 0.02--14 MeV energy range.}
  \begin{tabular}{lccccc}
  \hline
  Time interval & Spectral & $\Gamma$ & $E_{peak}$ & $\Gamma_2$ & $\chi^2/$dof \\
  (s) & model & & (keV) & & \\
  \hline
  Central part & Band & -0.95 (-0.10,+0.12)$^a$ & 660 (-161,+183) &  -2.34 (-1.23,+0.24) & 101/88 \\
  ($T_0$ to $T_0$+8.2) & CPL& -1.00 (-0.08,+0.09)& 770 (-123,+162)& -- & 105/89\\
   & PL& -1.46 (-0.02,+0.02)& -- & -- & 306/90\\
   & & & & & \\
  Main pulse & Band & -0.75 (-0.28,+0.40) & 307 (-101,+195) &  -2.10 (-0.59,+0.20) & 97/96 \\
  ($T_0$ to $T_0$+98) & CPL & -1.06 (-0.15,+0.18)& 563 (-157,+291) & -- & 101/97\\
   & PL& -1.51 (-0.04,+0.04)& -- & -- & 185/99 \\
   & & & & & \\
  Tail & PL$^b$ & -1.97 (-0.32,+0.39) & -- &  -- & 117/98 \\
  ($T_0$+98 to $T_0$+311) & & & & & \\
  \hline
  \end{tabular}
 \label{TableSpec}
\newline $^a${\footnotesize{All the quoted errors are at the 90\% confidence level.}}
\newline $^b${\footnotesize{In this region parameters of the CPL and Band models are not constrained.}}
\end{minipage}
\end{table*}

In the triggered mode, the burst time  histories are recorded by
the instrument in three energy bands: G1 (20--80~keV),
G2 (80--300~keV), and G3 (300--1200~keV). The measurement started
from 0.5~s before and up to 229.6~s after the spacecraft trigger, with a
time resolution varying from 2~ms up to 256~ms. The time history
of the burst is also available from the instrument's waiting-mode
data, which are recorded up to $T_0$+250~s in the same energy
bands with 2.944~s time resolution. A search for a possible
precursor to the burst was performed using this data set, with
no statistically significant positive result up to $T_0$-1500~s.

As observed by Konus-\textit{WIND}, the  burst light curve shows
a multi-peaked hard-spectrum pulse which starts at $\sim T_0$-25~s
and decays to $\sim T_0$+100~s (Fig.~\ref{FigLC}). Additional
information about the burst light curve after $T_0$+250~s can be
extracted from the multichannel spectra which are measured up to
$T_0$+490~s. The analysis of these data shows that the
$\gamma$-ray emission continues, mostly in the soft G1 band, up to
$\sim T_0$+310~s. The total burst duration $T_{100}$=246.1~s was
determined at the 5$\sigma$ level in the 20--1200~keV band. The
corresponding value of $T_{90}$ is 214$\pm$30~s and
$T_{50}=$48$\pm$4~s.

During the burst the instrument measured 64 multichannel energy
spectra covering two partially overlapping energy ranges: PHA1
(20--1200 keV) and PHA2 (0.4--14 MeV). The spectral analysis was
performed using XSPEC~V12.5~\citep{Arnaud1996} by applying three
spectral models. The first one is a simple power-law (PL): $f(E)
\propto E^{\Gamma}$ where $\Gamma$ is the power-law photon index.
The second model is a power-law with an exponential cut-off,
parametrized as $E_{peak}$: $f(E) \propto
E^{\Gamma} exp(-(2+\Gamma)E/E_{peak})$ where $E_{peak}$ is the
peak energy in the $\nu$F$_\nu$ spectrum. The third model is the
Band function~\citep{Band1993}: $f(E) \propto
E^{\Gamma}exp(-(2+\Gamma)E/E_{peak})$ for $E <
E_{peak}(\Gamma-\Gamma_2)/(2+\Gamma)$, and $f(E) \propto
E^{\Gamma_2}$ for $E \geq E_{peak}(\Gamma-\Gamma_2)/(2+\Gamma)$
where $\Gamma_2$ is the photon index in the higher energy band.

A summary of the Konus-\textit{WIND} spectral fits is presented
in Table~\ref{TableSpec}. The spectrum of the brightest part of the burst
(measured from $T_0$ to $T_0+$8.192~s) is best fitted in the
20--14000~keV energy range with the Band model ($\chi^2$=101/88~dof)
(see Fig.~\ref{FigSp15}). The CPL function yields a fit result of
similar quality ($\chi^2$=105/89~dof). For both mentioned models
relatively high values of $E_{peak}$~$\sim$700~keV are obtained,
placing this burst among the hardest of the Konus-\textit{WIND}
long GRBs. The time-integrated spectrum of the main part of the
GRB (measured from $T_0$ to $T_0$+98.3~s) is also best fitted by
the Band function with $E_{peak}=307^{+195}_{-101}$~keV
($\chi^2$=97/96~dof). Considering a spectrum of the final part of
the burst (from $T_0+$98.3 to $T_0+$311.3~s) the only model
for which the fit parameters are constrained is a simple
power-law function yielding the photon index $\Gamma$ close to -2
($\chi^2$=117/98~dof).

Based on the results of the spectral analysis the observed burst
energetics parameters such as energy fluence
(Table~\ref{TableFluence}) and peak energy flux are calculated.
The latter value reaches $F_{max} = (7.0 \pm 0.7) \times
10^{-6}$~erg~cm$^{-2}$~s$^{-1}$ in a 64~ms time interval starting
at $T_0$+1.1~s. The total energy fluence of the burst amounts to
$S = (6.5 \pm 1.1) \times 10^{-5}$~erg~cm$^{-2}$. Both values are
measured in the 20~keV -- 10~MeV energy range, standard for calculations
of the Konus-\textit{WIND} GRB energetics, and the quoted errors
are at the 90\% confidence level.

It is possible to consider the GRB time history in the three energy
bands (G1, G2, G3) as a series of three-channel energy spectra. In
this case the spectral variability of the burst can be studied on
a faster time scale \citep{Mazets2001,Ulanov2005}. We applied this
method to the GRB~051008 light curve in order to obtain the fine
temporal behaviour of the CPL spectral model parameters $E_{peak}$
and $\Gamma$ during the brightest phase of the burst. As a result
(Fig.~\ref{FigSpEvo}) clear evidence of spectral variability in
the form of a hardness-intensity correlation is obtained
(particularly between $E_{peak}$ and the light curves above
80~keV).

We also examined the spectral lag $\tau_{lag}$ using the cross-correlation
function (CCF) between the light curves in two energy bands
\citep{Norris2000,Band1997}. After calculating the CCF as a
function of the lag $\tau_{lag}$ we obtained the peak value of
$\tau_{lag}$ by fitting it with a fourth-degree polynomial. The
resulting values of $\tau_{lag}$ between the G1, G2, and G3 light
curves in different combinations are listed in Table~\ref{TableLag}.
As one can see none of these values differ from zero by more than
one standard deviation and thus a negligible spectral lag is evident
(see also Section 2.3).
This result places GRB~051008 among the 17\% of zero-lag events of
the Konus-\textit{WIND} sample of $\sim950$ long GRBs with
well-determined spectral lags\footnote{Svinkin et al., 2014, in preparation}.

%
\subsection{\textit{INTEGRAL} observations}
\begin{figure}
\centering
\includegraphics[width=84mm]{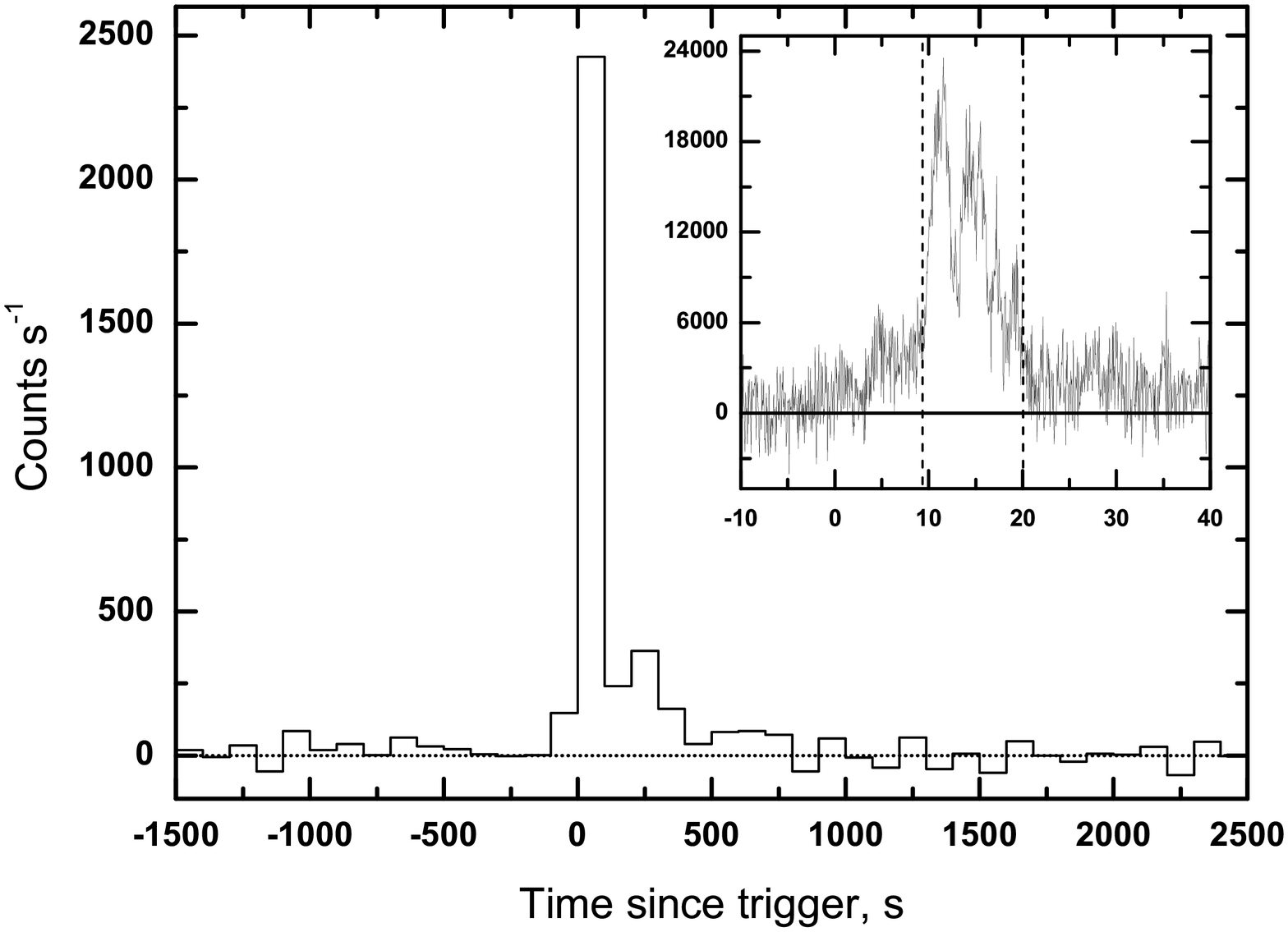} \\
a) \\
\includegraphics[width=84mm]{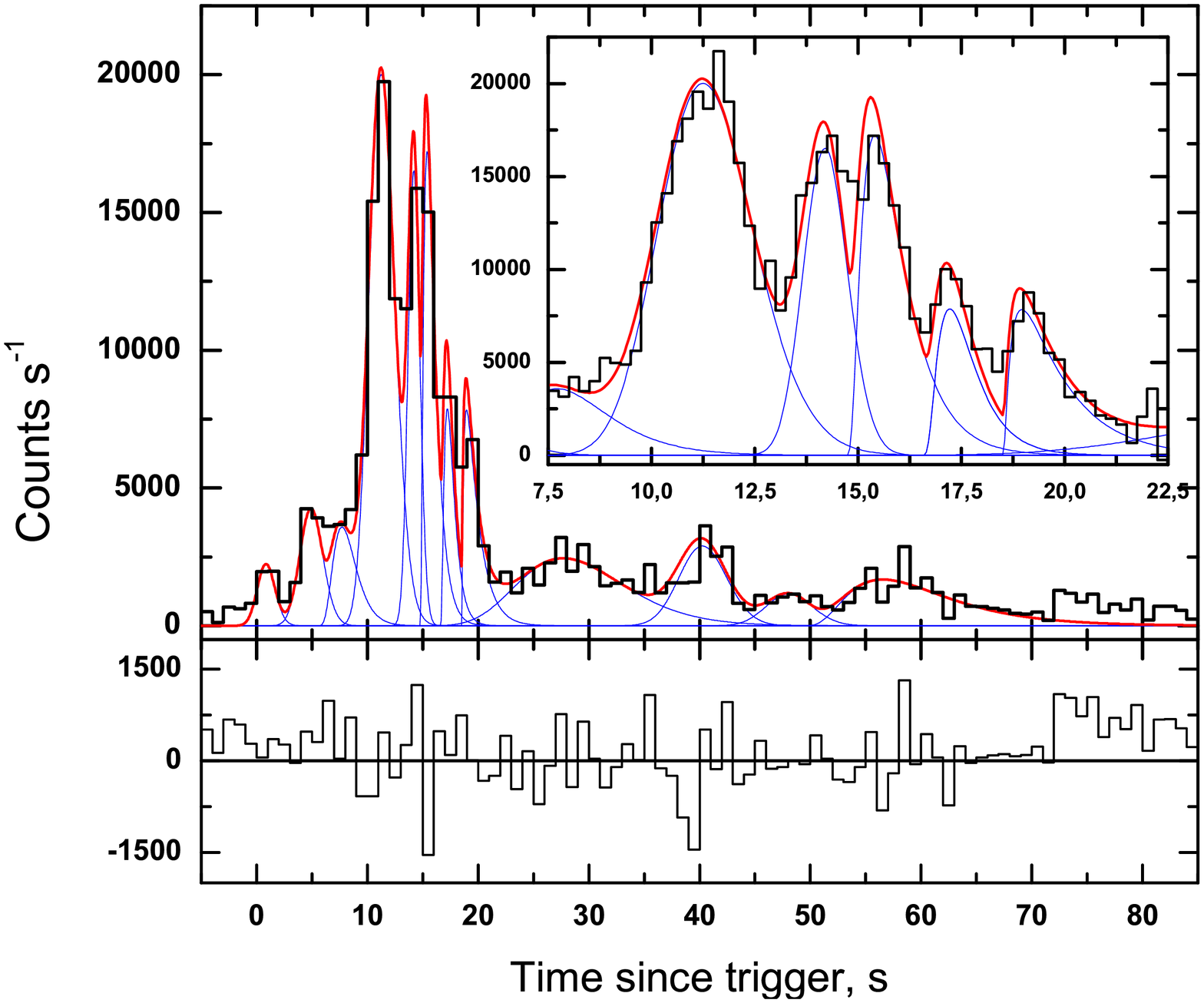} \\
b)\\
\caption{\textbf{a)} Background-subtracted light curve of
GRB~051008 obtained by the \textit{INTEGRAL}/SPI-ACS in the energy
band $\sim$ 80~keV -- 10~MeV. Here, the original light curve with a
time resolution of 50 ms has been re-binned into 100~s bins. Inset: the light
curve of the main peak of the burst; the dashed lines denote the
time interval for the PDS calculation, see Section~2.3. \textbf{b), top
panel} The light curve at 1 s time resolution fit with a number of single pulses;
\textbf{b), bottom panel} Residuals.}
\label{spi}
\end{figure}

The burst was detected by the SPI-ACS detector on-board the
\textit{INTEGRAL}  observatory at 16:33:12
UT\footnote{{\scriptsize{\tt
http://www.isdc.unige.ch/\textit{INTEGRAL}/ibas/\newline
cgi-bin/ibas\underline{ }acs\underline{ }web.cgi/?trigger=2005-10-08T16-33-12.1588-07660-00007-0}}}.
The burst position was $\theta$ = 86 degrees off the axis of the aperture telescopes.
The SPI-ACS continuously records count rates in a single wide
energy band $\sim$ (80--10000) keV with a time resolution of 50 ms
\citep{spi}. Owing to the large size of the SPI-ACS detector
(geometrical surface area is $5250~cm^{2}$ at the $\theta=90$)
significant ($3\sigma$) emission was detected up to 800 s after
the burst trigger (Fig.\ref{spi}). We again searched for a
possible precursor but did not find one in the range of
$T_0-1500$~s. The $T_{90}$ in the energy band of SPI-ACS is $535
\pm 40$~s.

Using the original time resolution we searched for a possible
periodicity of the whole light curve. No significant periodicity
was found on time scales from 0.1 to 20 s. We also investigated
the power spectrum of the main peak of the  burst in the time
interval [10--20] s after trigger (see inset in Fig.\ref{spi}).
The power density spectrum (PDS) can be fitted by a single
power-law with index of $-2.06 \pm 0.09$ up to 7~Hz. The
power-law index significantly differs from the index $\sim 1.6$ of
averaged PDS obtained for long GRBs \citep{belobor,poz}.

GRB~051008 is a good example of a multi-peak event. Using SPI-ACS
data and technique developed early \citep{minaev} we found at least 
10 separated pulses which fit the light curve fairly well (Fig.\ref{spi}b). 
We cannot investigate the spectral lag of the separate pulses due to heavy 
overlap of the pulses. It was found the existence of the lag vs. pulse 
duration correlation \citep{hakkila,minaev_eas,minaev}, i.e. 
the shorter pulse, the smaller spectral lag corresponds to the pulse. 
It is multi-peak structure of short duration pulses composing the main 
body of a burst light curve could be the underlying nature of a negligible 
lag of the GRB~051008 as a whole (see Table~\ref{TableLag}, Fig.~\ref{spi}).

Since the lag of the whole burst is not an additive function of the lag 
of separate pulses even superposition of pulses with positive lags may 
result in a negative lag of a complex of superimposed pulses \citep{minaev}.
This is another plausible reason of the negligible lag of the GRB~051008.

%
\subsection{\textit{Swift}/XRT observations}

The XRT X-ray telescope \citep{xrt} on-board the \textit{Swift}\,
space observatory began observing the region of the burst at 17:23:52
UT, i.e., 50 minutes after the BAT trigger. The XRT detected the fading
X-ray source \citep{per} which was identified as the X-ray afterglow
of the burst.
\citet{butler} provided an on-ground SDSS-enhanced
position of the X-ray afterglow at RA $= 13^h31^m29\fs55$, Dec.
$= +42\degr05\arcmin53\farcs3$ with an accuracy of $1\farcs2$
(90\% confidence level). For additional information on the X-ray afterglow
analysis see \citet{margutti}.

\begin{figure}
\centering
\includegraphics[width=84mm]{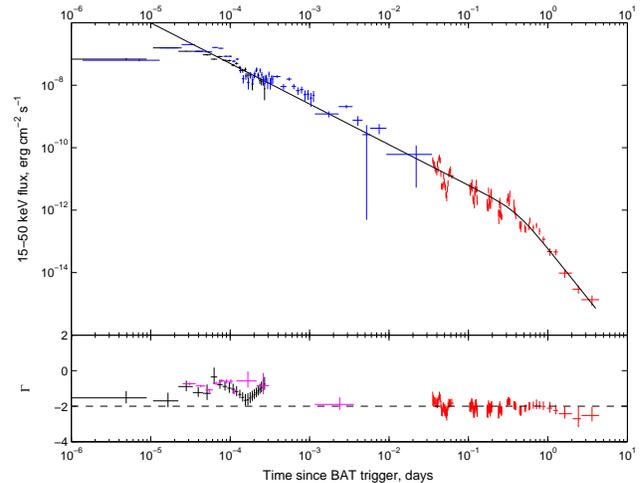}
\caption{\textbf{(top)} The 15--50~keV X-ray light curve of GRB~051008 based on
the data from the \textit{Swift}/BAT (black crosses), \textit{Swift}/XRT
(red crosses), and \textit{INTEGRAL}/SPI--ACS (blue crosses). The black
solid line represents the best fit to the data using a broken power-law
(1); \textbf{(bottom)} photon indices $\Gamma$ of the X-ray spectrum obtained by
\textit{Swift}/BAT (black crosses), \textit{Swift}/XRT (red crosses),
and Konus-\textit{WIND} (magenta crosses).}
\label{XRT}
\end{figure}

Fig.~\ref{XRT} shows the flux light curve in the 15--50 keV
energy range created using BAT, XRT\footnote{{\scriptsize{\tt
http://www.swift.ac.uk/burst\underline{
 }analyser/00158855/\newline SHOWbat\underline{ }xrt\underline{
 }lightcurves/TIMEDEL1/\newline bat\underline{ }xrt\underline{
 }flux\underline{ }with\underline{ }gamma\underline{
 }BATBAND\underline{ }BATTIMEDEL1.qdp.gz}}}, and SPI-ACS data. The
flux light curve can be created from a count-rate light curve
using a conversion factor computed from spectral and absorption
models and instrumental response matrices. The procedure is not
unreasonable if the photon index ($\Gamma$) varies smoothly with
time \citep[for more details see][]{evans2}. Indeed the photon
index (Fig.~\ref{XRT}, bottom panel) is seen to be monotonously
decreasing from the end of the BAT data toward XRT via the
Konus-\textit{WIND} data in between.

Since the SPI-ACS experiment has no spectral channels we
converted the SPI-ACS count-rate light curve into a flux light
curve in the energy range 15--50~keV in two steps. (1) Using the
Konus-\textit{WIND} CPL spectral model we calculated the fluence
of the main activity episode (-25~s -- 98~s) in the 80~keV -
10~MeV energy range (i.e., the SPI-ACS energy band), and we found a
conversion coefficient (at the same time interval of the SPI-ACS)
which is equal to $(2.5 \pm 0.7)\times 10^{-10}$ erg/cm$^2$ per count.
(2) Using the same CPL spectral model we converted the SPI-ACS
fluence into the 15--50~keV energy range.

The resulting BAT+SPI-ACS+XRT flux light curve was fitted by a
broken power-law \citep[see, e.g.,][]{beuermann}:
\begin{equation}
F = F_0\left[ \left( \frac{t}{t_{b}}\right) ^{\alpha_1w} + \left( \frac{t}{t_{b}}\right) ^{\alpha_2w} \right] ^{-1/w},
\end{equation}
where $\alpha_1$ and $\alpha_2$ are the initial and final flux
decay indices; $t_{b}$ is the time corresponding to the break in
the light curve; $w$ is the parameter responsible for the
sharpness of the break, and $F_0$ is the normalization
coefficient. Figure~\ref{XRT} shows the light curve with the fit.
The fit yields the following parameter values: $t_{b} = 0.41 \pm
0.14$~days, $\alpha_1 = 1.30 \pm 0.18$, and $\alpha_2 = 3.18 \pm
0.20$. Among tested discrete values of break sharpness from 1 to 5
the best-fitted $w$ equals 2.

The $t_b = 0.41 \pm 0.14$ days is somewhat larger than the one
obtained in~\citet{racusin} ($t_b = 0.19 \pm 0.04$ days). It could
be due to different light-curve models being used, but primarily
due to adding \textit{Swift}/BAT and \textit{INTERGRAL}/SPI-ACS
data to the fit. Indeed, if we fit only the \textit{Swift}/XRT
data~\citep{meastbul}, we find this parameter to be in a good agreement
with the results of~\citet{racusin}. It is also evident that the
afterglow light curve of GRB~051008 belongs to a less frequent class
of plateau-less XRT light curves and may represent only phases III
and IV of the canonical X-ray afterglow~\citep{racusin,zhang2006,nousek,liang3}.
%
\subsection{Optical and NIR observations}

\begin{figure}
\centering
\includegraphics[width=84mm]{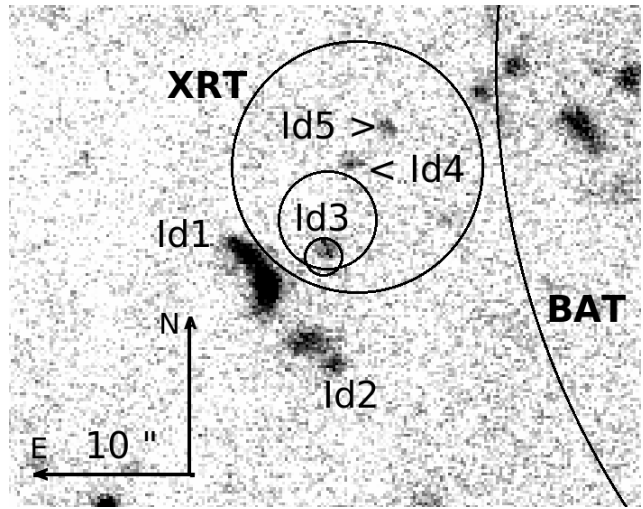}
\caption{The localization of GRB~051008 and its immediate
neighbourhood. The image was taken 255 days after the GRB in the
$R$-band filter (the limiting magnitude is $25\fm7$) with a 4800~s
exposure, using the NOT. The burst error
circles determined from the \textit{Swift}/BAT\, and the
\textit{Swift}/XRT\, telescopes data are shown with solid lines.
An arc at the right side of the image is a part of BAT error
circle. Three small circles denote different XRT error boxes. The
position of the X-ray afterglow of the burst lies outside the
refined BAT error circle. The smallest error box depicts the
\textit{Swift}/XRT refined error circle at (J2000) RA $=
13^h31^m29\fs55$, Dec. $= +42\degr05\arcmin53\farcs3$ with a
radius of $1\farcs2$ \citep{butler}. The Id1--5 markers indicate
the sources studied in this paper. The coordinates of the Id3
source are  RA $= 13^h31^m29\fs510$, Dec. $=
+42\degr05\arcmin53\farcs67$ (J2000) with an error of $0\farcs08$
in both right ascension and declination (statistical error only).
The image scale and seeing are equal to $0\farcs19$ per pixel and
$0\farcs9$, respectively.}
\label{NOT}
\end{figure}

The Crimean Astrophysical Observatory started observations of the burst
error box with the 2.6-m Shajn telescope 32 minutes after the burst
trigger \citep{ztsh}. Close to the initial XRT error circle a source
(Id1 in Fig. \ref{NOT}) was found, which however was later discarded
as the OA of GRB~051008~\citep{crao}. The limiting $R$ magnitude of
the co-added frame with mean observing time 17:28:14 UT is $23\fm3$.

The UVOT optical telescope of the \textit{Swift}\, space observatory
started to monitor the region of GRB~051008 at 17:23:49 UT, 50 minutes
after the burst trigger~\citep{roming}. The telescope took a 200-s $v$-band
image, no afterglow was found at a 5-$\sigma$ significance level for
$18\fm2$ \citep{breev}. The optical observations of the burst region
were highly hindered by the presence of the bright $R \sim 5\fm5$ star
HR 5096 $2\farcm5$ from the XRT position.

\textit{Swift}/UVOT continued observations of the GRB~051008
region up to Oct.14 using the ultraviolet filters $u, uvw1, uvm2$
and $uvw2$, with a total exposure of about 11000 seconds in each
filter, and the $v$ filter with a total exposure of more than 120000 s.
Neither OT nor the host galaxy of the GRB was detected in stacked
images in any UVOT filter, particularly up to $uvw2 > 23.2$.

Observations of the field of GRB~051008 were obtained with the
1.34m Schmidt telescope of the Th\"uringer Landessternwarte
Tautenburg as soon as dusk fell, beginning 0.8 hours after the GRB
trigger \citep[see][]{klose,fer,fer2}. Initial observations were
obtained in the $I_C$ band, lengthening the exposure times
($2\times10$ s, $6\times20$ s, $7\times120$ s) in order to
accommodate for the decrease in sky background. These observations
were followed by sets of $6\times120$ s each in $R_C$ and $V$. The
halo of the bright nearby star influences the GRB position only
in the $R_C$ image, reducing the limiting magnitude.

We obtained further imaging with the TLS
Schmidt $\sim20$ days after the GRB, around the expected peak-time
of a low-$z$ GRB supernova \citep{zeh,fer2}. We obtained $30\times60$ s
exposures in the $I_C$ band, and then $3\times30\times60$ s
exposures in the $R_C$ band in three consecutive nights. These
observations were all obtained at high airmass (1.8 -- 3.0). For
the 90 minute stacked $R_C$ image, we carefully modelled and
subtracted the flux of the nearby bright star. While strong
residuals remain within $20^{\prime\prime}$ of the star's
position, the GRB location is far enough away to have a smooth
background. We clearly detect the galaxy Id1 and faintly detect
the galaxy Id2 next to it. At the XRT position of the X-ray
afterglow (and the host galaxy Id3 underneath it), no source is
detected to $R_C>22.7$. The $I_C$ upper limit is significantly
less deep and not constraining.

All TLS data were calibrated against two bright but non-saturated
SDSS DR8 stars (stars 6 and 7 in Table \ref{refstars}), as the
other stars were in most cases not even detected. SDSS AB
magnitudes were transformed into Johnsons-Cousins Vega magnitudes
analogue to the other calibration stars.

On April 28, 2006 a source candidate for the host galaxy of
GRB~051008 was found as a result of observations made with the
Shajn telescope of the Crimean Astrophysical Observatory (see Fig.\ref{NOT}, 
source Id3). The source was further observed in 2006--2009 with the Nordic Optical
Telescope (NOT), the AZT-22 telescope of the Maidanak Observatory,
and the Keck telescope. These observations allowed us to take
images of the burst region in the $Bg^{\prime}VRR_CI_C$ filters
and in the interference filter $i$. In February 2008 and in February--March
2012 we obtained additional optical observations of the field
using the Low Resolution Imaging Spectrometer (LRIS) on Keck. The
series of frames were taken in the $UBIZ$ filters with total exposure
times of 930, 930, 810, and 720 seconds, respectively. All
preliminary reductions were performed using private IDL codes. The
best limiting magnitude was obtained in the filter $g^{\prime}>
27\fm2$, for the other limiting magnitudes see Table~\ref{log}.

We acquired deep infrared imaging at the position of GRB~051008
using the Near-Infrared Imager (NIRI) on Gemini-North on the night
of 2010 June 23, starting at 07:15 UT. 27 exposures of 60 seconds
each were acquired on the field in the $K^{\prime}$ filter, which
we reduced and stacked using the routines in the Gemini IRAF
package. Unfortunately, there is only a single faint star (2MASS
13313142+4206470) in the FOV of the combined image, which a series
of late-time PAIRITEL images suggests may be variable. Instead,
since the night was photometric, we calibrate the host photometry
using an observation of the IR standard star FS 29 taken later in the
night of Gemini observations.

Table~\ref{log} presents the log of our observations of the
GRB~051008 field. Fig.~\ref{NOT} displays the location of
GRB~051008, indicating the supposed host galaxy, and the XRT and
BAT error circles. Galaxies Id1, Id2, Id4, and Id5 lie within a 10
arcsecond radius circle around the probable host galaxy of
GRB~051008 (Id3) and we will also study these galaxies below.
Table~\ref{phot} gives the results of the photometric reduction of
the observations of this region.
%
\begin{table*}
\centering
 \begin{minipage}{140mm}
  \caption{Log of optical and NIR observations of the GRB~051008 location.
  The magnitudes are not corrected for Galactic reddening.}
  \begin{tabular}{r|c|c|c|c|c|c}
  \hline
  Date  & $t-t_0$, & Telescope & Exposure, & FWHM, & Filter & Upper limit, \\
   & days & & s &  $\arcsec$ & & at the $3 \sigma$ level \\
  \hline\
  8.10.2005  & 0.034 & TLS & 20 & 1.3 & $I_C$ & 16.0 \\
  8.10.2005  & 0.035 & ZTSh & 2200 & 2.1 & $R_C$ & 23.3 \\
  8.10.2005  & 0.041 & TLS & 120 & 1.3 & $I_C$ & 17.5 \\
  8.10.2005  & 0.049 & TLS & 840 & 1.4 & $I_C$ & 19.3 \\
  8.10.2005  & 0.062 & TLS & 720 & 1.4 & $R_C$ & 20.0 \\
  8.10.2005  & 0.074 & TLS & 720 & 1.6 & $V$ & 21.5 \\
  8-10.10.2005  & 0.819 & UVOT & 11474 & 2.5 & $uvw2$ & 23.2 \\
  8-10.10.2005  & 0.824 & UVOT & 10822 & 2.5 & $uvw1$ & 22.5 \\
  8-10.10.2005  & 0.845 & UVOT & 11887 & 2.5 & $uvm2$ & 22.7 \\
  8-10.10.2005  & 0.887 & UVOT & 11181 & 2.5 & $u$ & 22.5 \\
  10-14.10.2005 & 4.128 & UVOT & 125336 & 2.5 & $v$ & 22.8 \\
  27.10.2005 & 19.46 & TLS & 1800 & 1.7 & $I_C$ & 20.0 \\
  29.10.2005 & 21.47 & TLS & 5400 & 2.5 & $R_C$ & 22.7 \\
  28.04.2006 & 202.3 & ZTSh & 4260 & 1.9 & $R_C$ & 24.1 \\
  20.06.2006 & 255.3 & NOT & 4800 & 0.9 & $R$ & 25.7 \\
  30.06.2006 & 265.2 & NOT & 4800 & 1.2 & $V$ & 25.7 \\
  25.07.2006 & 290.1 & ZTSh & 2760 & 1.5 & $R_C$ & 23.5 \\
  12.08.2006 & 308.2 & NOT & 3300 & 0.7 & $B$ & 25.9 \\
  20.08.2006 & 316.2 & NOT & 4200 & 0.9 & $i$ & 22.5 \\
  16.09.2007 & 709.9 & AZT-22 & 12120 & 2.0 & $I_C$ & 22.5 \\
  12.02.2008 & 856.9 & Keck I & 780 & 0.85 & $R_C$ & 26.1 \\
  12.02.2008 & 856.9 & Keck I & 960 & 0.85 & $g^{\prime}$ & 27.2 \\
  20.05.2009 & 1289.9 & NOT & 8400 & 1.1 & $i$ & 23.9 \\
  24.06.2010 & 1719.0 & Gemini & 1620 & 0.5 & $K^{\prime}$ & 23.0 \\
  20.02.2012 & 2325.7 & Keck I & 930 & 1.8 & $B$ & 25.8 \\
  20.02.2012 & 2325.7 & Keck I & 810 & 1.8 & $I$ & 25.2 \\
  15.03.2012 & 2349.6 & Keck I & 930 & 1.1 & $U$ & 25.4 \\
  15.03.2012 & 2349.6 & Keck I & 720 & 1.1 & $Z$ & 25.5 \\
  \hline
  \end{tabular}
 \label{log}
 \end{minipage}
\end{table*}
%
\begin{table*}
\centering
 \begin{minipage}{140mm}
  \caption{Photometric magnitudes (Vega) of the Id1--5 sources based
  on the observations made with the \textit{Swift}/UVOT and TLS in
  October 2005, ZTSh in April 2006, the NOT in June--August 2006
  and May 2009, Keck in 2008-2012 and Gemini in 2010. The magnitudes
  displayed in the table are in the Vega photometric system and have
  been dereddened for Galactic extinction \citep{shleg}.}
  \begin{tabular}{c|c|c|c|c|c|c}
  \hline
  Filter & Id1, $^m$   & Id2, $^m$        & {\bf Id3,} $\mathbf{^m}$  & Id4, $^m$  & Id5, $^m$ & Instrument \\
   & & & & &   \\
  \hline
  $uvw2$ & 21.69 (0.18) & 22.29 (0.26) & $\mathbf{> 23.3}$ & $> 23.3$ & $> 23.3$ & \textit{Swift}/UVOT \\
  $uvm2$ & 21.17 (0.16) & 22.66 (0.33) & $\mathbf{> 22.8}$ & $> 22.8$ & $> 22.8$ & \textit{Swift}/UVOT \\
  $uvw1$ & 21.31 (0.19) & 22.30 (0.39) & $\mathbf{> 22.6}$ & $> 22.6$ & $> 22.6$ & \textit{Swift}/UVOT\\
  $u$ & 21.32 (0.16) & $> 22.6$ & $\mathbf{> 22.6}$ & $> 22.6$ & $> 22.6$ & \textit{Swift}/UVOT\\
  $U$ & 21.30 (0.04) & 22.92 (0.07) & \bf{25.32 (0.23)} & 25.13 (0.19) & 25.46 (0.28) & Keck~I/LRIS \\
  $B$ & 21.80 (0.20) & 23.63 (0.21) & \bf{25.20 (0.23)} & 25.28 (0.27) & 25.70 (0.26) & NOT/ALFOSC \\
  $B$ & 22.12 (0.03) & 23.44 (0.07) & \bf{25.27 (0.13)} & 25.22 (0.13) & 25.67 (0.19) & Keck~I/LRIS \\
  $g^{\prime}$ & 21.77 (0.05) & 23.49 (0.07) & \bf{24.57 (0.07)} & 24.94 (0.13) & 25.19 (0.19) & Keck~I/LRIS \\
  $v$ & 22.27 (0.29) & $> 22.8$ & $\mathbf{> 22.8}$ & $> 22.8$ & $> 22.8$ & \textit{Swift}/UVOT\\
  $V$ & 21.50 (0.13) & 22.93 (0.14) & \bf{24.45 (0.13)} & 24.75 (0.18) & 24.91 (0.19) & NOT/ALFOSC \\
  $R$ & 21.15 (0.04) & 22.30 (0.09) & \bf{24.07 (0.11)} & 24.16 (0.12) & 24.38 (0.11) & NOT/ALFOSC \\
  $R$ & 21.35 (0.08) & 22.42 (0.13) & \bf{24.01 (0.24)} & $> 24.1$ & $> 24.1$ & ZTSh \\
  $R_C$ & 21.22 (0.05) & 22.39 (0.07) & \bf{24.06 (0.10)} & 24.13 (0.13) & 24.33 (0.13) & Keck~I/LRIS \\
  $R_C$ & 21.18 (0.08) & 22.28 (0.26) & $\mathbf{> 22.7}$ & $> 22.7$ & $> 22.7$ & TLS \\
  $I$ & 20.23 (0.03) & 21.33 (0.02) & \bf{23.88 (0.11)} & 23.90 (0.11) & 24.14 (0.14) & Keck~I/LRIS \\
  $i$ & 20.37 (0.05) & 21.53 (0.06) & \bf{24.01 (0.15)} & 23.81 (0.17) & 23.98 (0.16) & NOT/ALFOSC \\
  $Z$ & 20.29 (0.08) & 21.31 (0.09) & \bf{23.80 (0.13)} & 23.78 (0.18) & 23.93 (0.18) & Keck~I/LRIS \\
  $K^{\prime}$ & 18.49 (0.10) & 19.18 (0.11) & \bf{22.86 (0.29)} & $> 23.0$  & $> 23.0$ & Gemini/NIRI \\
  \hline
  \end{tabular}
  \label{phot}
 \end{minipage}
\end{table*}
%
\begin{table*}
 \centering
 \begin{minipage}{140mm}
  \caption{Reference stars used for the source photometry. The data are
  adopted from the SDSS-DR7 catalogue \citep{sdss}. The magnitudes are
  in the AB photometric system.}
  \begin{tabular}{c|c|c|c|c|c|c|c}
  \hline
  Star's ID & RA, J2000 & Dec, J2000 & $u$ & $g$ & $r$ & $i$ & $z$ \\
  \hline
  J133138.56+420443.5 & 13 31 38.57 & $+42$ 04 43.6 & 22.110 & 20.507 & 20.028 & 19.826 & 19.755 \\
  J133137.89+420705.4 & 13 31 37.89 & $+42$ 07 05.5 & 20.073 & 19.010 & 18.831 & 18.742 & 18.694 \\
  J133141.93+420803.0 & 13 31 41.93 & $+42$ 08 03.1 & 23.900 & 22.618 & 21.203 & 19.684 & 18.925 \\
  J133146.39+420825.9 & 13 31 46.39 & $+42$ 08 25.9 & 21.217 & 19.594 & 19.097 & 18.827 & 18.755 \\
  J133145.90+420730.4 & 13 31 45.91 & $+42$ 07 30.4 & 22.789 & 20.456 & 19.836 & 19.586 & 19.367 \\
  J133133.58+420401.4 & 13 31 33.58 & $+42$ 04 01.5 & 15.456 & 13.973 & 13.497 & 13.769 & 13.286 \\
  J133153.54+420304.7 & 13 31 53.54 & $+42$ 03 04.8 & 19.365 & 16.792 & 15.435 & 14.779 & 14.414 \\
  \hline
  \end{tabular}
 \label{refstars}
 \end{minipage}
\end{table*}

All astrometric solutions were obtained using the APEX astrometric
code \citep{apex}. All astrometric errors are statistical only if
not stated otherwise. The reference catalogue for the
astrometric reduction is USNO-B1.0. All photometric calibrations
for the data employed in this paper were performed using the
method of relative aperture photometry via the APPHOT procedure
of the IRAF software package\footnote{{\scriptsize{\tt
http://iraf.noao.edu/}}} except for the \textit{Swift}/UVOT data which
were reduced using NASA's HEASARC software package
HEAsoft\footnote{{\scriptsize{\tt
http://heasarc.nasa.gov/docs/software/lheasoft/}}}. The Id1 and
Id2 sources have complex shapes and using the APPHOT procedure we
therefore applied the apertures shaped as irregular closed
polygons inscribed in the isophotes. For the photometric
calibration of the $Bg^{\prime}VRI$ images we used seven stars
from the SDSS-DR7 catalogue~\citep{sdss} (see
Table~\ref{refstars})\footnote{{\scriptsize Magnitudes are
converted from the \textit{ugriz} into the \textit{BVRI} system
using Lupton's transformations {\tt
http://www.sdss.org/dr7/algorithms/sdssUBVRITransform.html}}}.
The transmission curves of the Keck $U$
and $Z$ filters do not match exactly the SDSS $u$ and $z$ bands.
The $U$ and $Z$ images were calibrated using the relative
magnitude averaged by 14 non-overexposed stars of the field.
The stars were chosen with the aid of an on-line software
package which we use to search for secondary photometric
standards in the FOV of GRBs from automatic GCN notices
(BACODINEs).\footnote{{\scriptsize
Pozanenko et al., in preparation.}} To process \textit{Swift}/UVOT
data we used the task UVOTIMSUM of the HEAsoft package to combine
all frames for each filter ($u, uvw1, uvm2, uvw2$ and $v$) and the
task UVOTSOURCE to obtain the magnitudes of the sources Id1--5.
The calibration and upper limit estimates were done using the
HEASARC CALibration Data Base (CALDB\footnote{{\scriptsize {\tt
http://heasarc.gsfc.nasa.gov/FTP/caldb}}}). For the galaxies Id3--5
we used circular apertures with a fixed radius of $2\farcs2$.
For the Id1 and Id2 galaxies we increased this radius to $5\farcs0$.
%
\subsection{Spectroscopy of the host galaxy}

We obtained a longslit spectrum of the host galaxy of GRB~051008
on the night of 2009 June 25 (UT) using the Low-Resolution Imaging
Spectrometer \citep[LRIS,][]{oke} on the Keck I 10-meter
telescope. Three exposures of 900 seconds each were acquired on
the source at a position angle of 202 degrees, using the D560
dichroic, 600-line grism (blue), and 400-line grating (red).  The
target was acquired independently before each of the three 900~s
exposures and no additional dithering was performed, so the
exposures are effectively undithered aside from small differences
of less than 1 arcsecond.
No obvious line features are identified in the sky-subtracted 2D spectra.
%
\subsection{Radio observations}

GRB~051008 was observed by the  Very Large Array. No radio
afterglow was detected. An 8.5 GHz flux upper limit of 0.12~mJy
was obtained for the burst location region on Oct. 8, 2005, at
20:30:24 UT, i.e., four hours after the trigger~(\citealt{cam}, see
also \citealt{chandra}). Further observations carried out 2 days
after the trigger on Oct. 10, 18:57:36 UT at the same frequency
obtained a flux upper limit of 0.11 mJy\footnote{{\scriptsize{\tt
http://www.aoc.nrao.edu/$\sim$dfrail/grb051008.dat}}}.

The position was also re-observed on June, 11, 2012 with the upgraded
Karl G. Jansky Very Large Array as part of the host-galaxy study
of \citet{doubleperley}. No source was detected consistent with
the XRT position with a 2$\sigma$ upper limit of 0.014 mJy at 5.23 GHz.
%
\section{PROPERTIES OF THE HOST GALAXY}

The coordinates of the Id3 source (Fig. \ref{NOT}) are RA
$=13^h31^m29\fs510$, Dec. $= +42\degr05\arcmin53\farcs67$ (J2000),
accurate to within $0\farcs08$ in both coordinates (statistical
error). The source lies inside the final XRT $1\farcs2$-error box
corrected using the joint observations of the source field
performed by the XRT and the centroids of corresponding optical
sources in the SDSS-DR5 catalogue~\citep{butler}. However,
this match may be accidental. \citet{bloom} reported a formula for
computing the probability of finding a galaxy inside an arbitrary
sky area as a function of the galaxy magnitude. For GRB~051008 it
yields a probability of about $1.96$\% for an accidental location
of the Id3 galaxy inside the XRT error circle.

In a circle with centre at Id3
and radius of $10\arcsec$ there are 5 galaxies.
One can assume that these galaxies compose a group.
We will investigate it below.

\begin{figure*}
\centering
\begin{minipage}{180mm}
\begin{tabular}{cc}
\includegraphics[width=85mm,bb=1 147 718 694,clip=true]{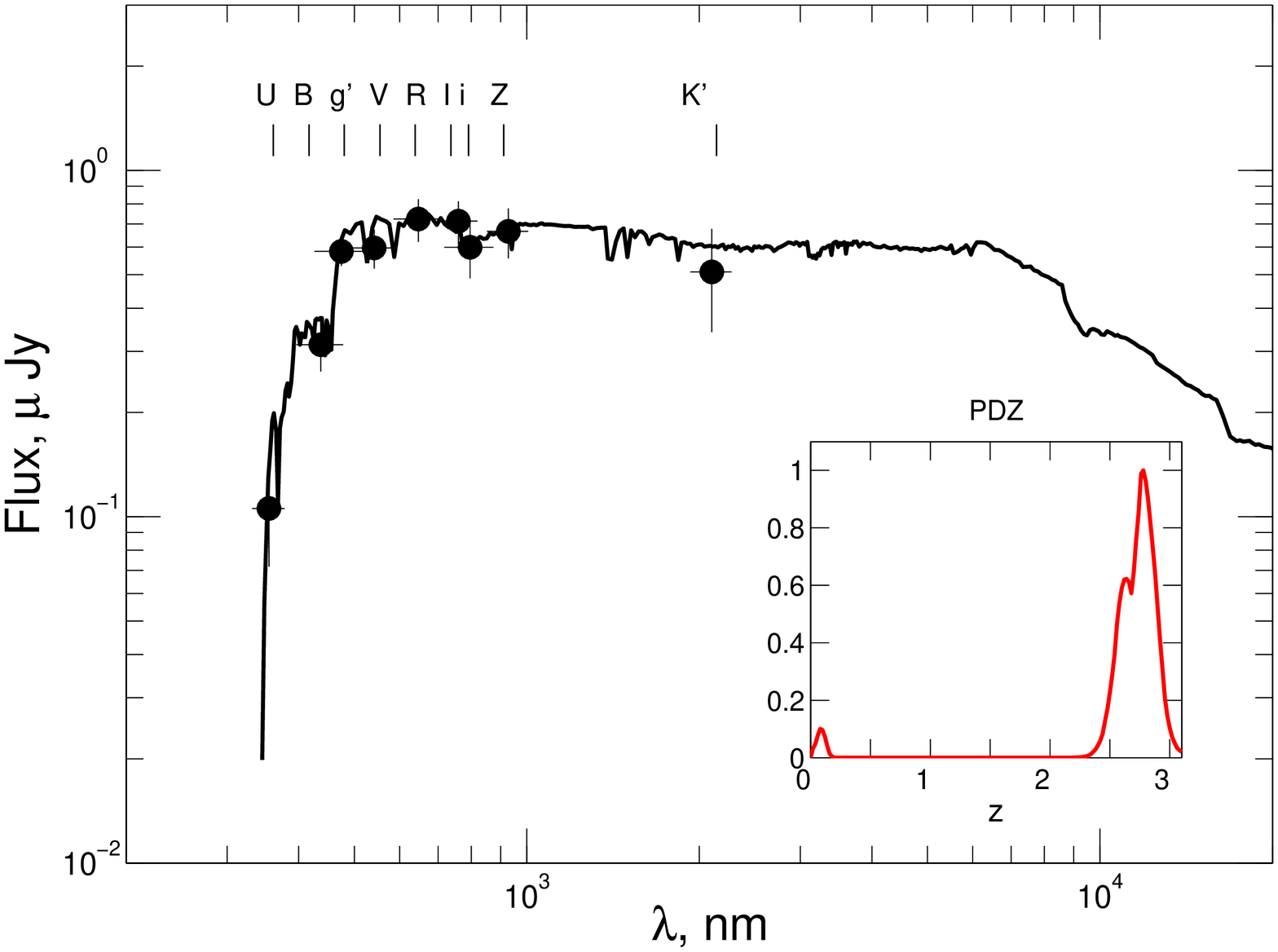} &
\includegraphics[width=85mm,bb=1 147 718 694,clip=true]{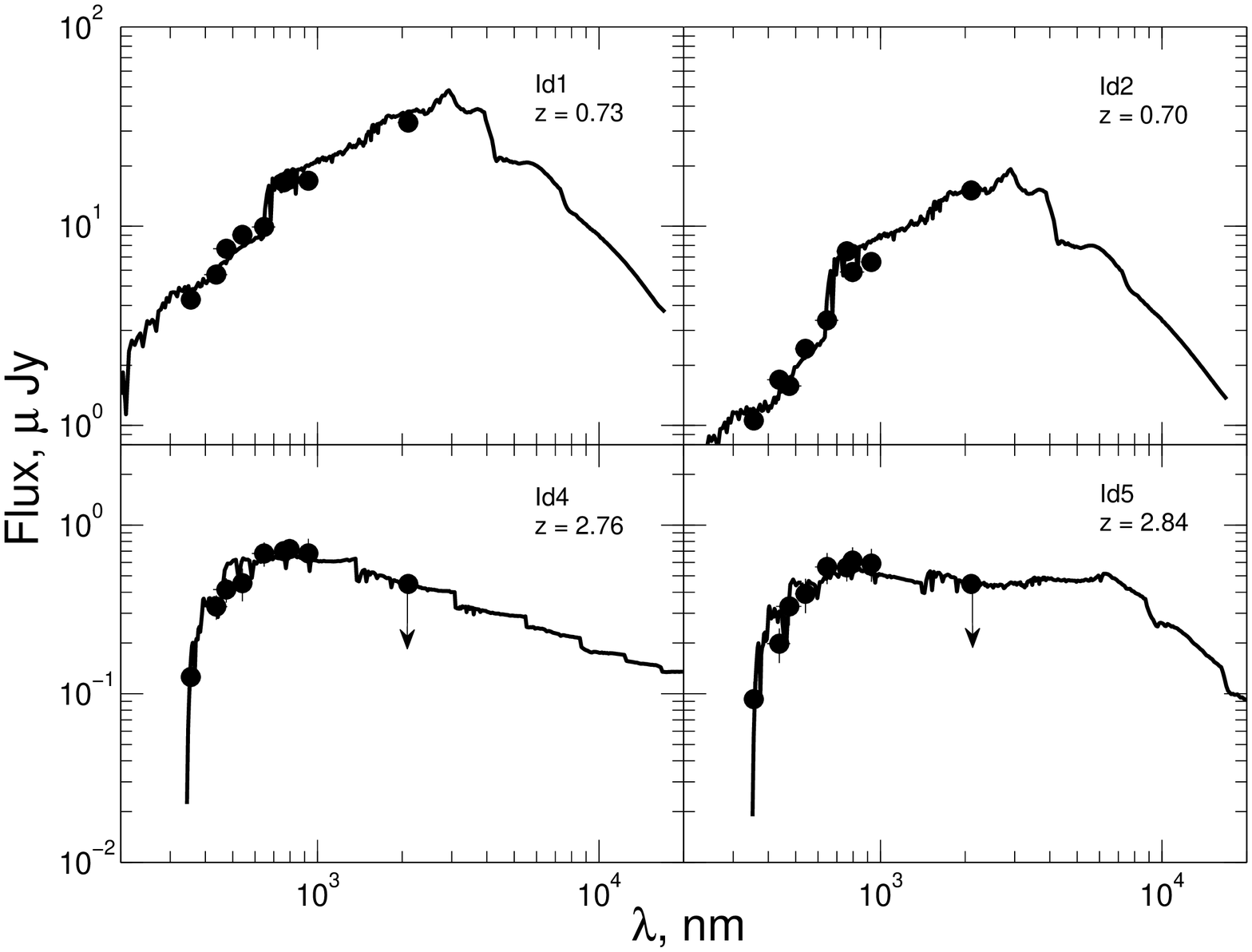} \\
a) & b) \\
\end{tabular}
\caption{\textbf{a)} Spectral Energy Distribution (SED) of the galaxy Id3 in the
observer frame. The best-fitted SED type obtained by \textit{Le Phare}
(line) is the SED of a starburst galaxy. Observed flux in
$UBg^{\prime}VRiIZK^{\prime}$ filters is shown by black circles.
The associated Probability Distribution Function is shown in the inset.
\textbf{b)} The best-fitted SEDs of the galaxies Id1, Id2, Id4 and Id5 obtained
by \textit{Le Phare} with observational data marked as in the figure a).}
\label{sed}
\end{minipage}
\end{figure*}

We used the \textit{Le Phare} software
package~\citep{lephar1,lephar2} to estimate the redshifts of the
galaxies Id1--5 from the photometric data obtained with the NOT,
Keck I, and Gemini North.
We found a photometric redshift of $z = 2.77_{-0.20}^{+0.15}$
(95\% confidence level) for the host galaxy. An independent analysis of
the redshift based only on the Keck observations shows $z =
2.90_{-0.16}^{+0.29}$ at the same confidence level \citep{perley2013}.
Table~\ref{z} lists the estimated redshift values and other
parameters of the galaxies Id1--5 including those determined using
\textit{Le Phare}: the most probable galaxy SED type, the best-fitted
extinction law, internal bulk extinction in the galaxy, and age
of the dominant stellar population of the galaxy. We used the
PEGASE2 population synthesis models library \citep{pegase} to obtain the
best-fitted SED, the redshift, and the other required parameters.
For all 5 galaxies we tried 3 different reddening laws: LMC
\citep{lmc}, SMC \citep{smc}, and the reddening law for
starburst galaxies \citep{calzetti}. In case of the galaxy Id3
the best-fit is the LMC reddening law. The best-fit reddening
laws for the galaxies Id1,2,4,5 are listed in
Table~\ref{z}. Figure~\ref{sed}a shows the best-fit SED of the
galaxy Id3. It reveals that the burst occurred in a moderately
absorbed LBG. It is one of the few cases where
a LBG hosts a GRB \citep[see also][]{971214,021004,malesani}.

The redshift of $z=2.77$ corresponds to a luminosity distance of
23.5 Gpc and a distance modulus of 46\fm8, one arcsecond
corresponds to 8.0 proper kpc. For the galaxies Id3--Id5 the
angular size is estimated as the full width at half maximum of the
point spread function during the observations made on February 12,
2008 (see Table \ref{z}), which is equal to 0\farcs9.

It is worth noting that the galaxies Id1 and Id2
have redshifts of $z_{Id1} = 0.73^{+0.10}_{-0.08}$ and
$z_{Id2} = 0.70^{+0.11}_{-0.08}$ (95\% confidence level) with
corresponding probability of $>99.99\%$. These redshifts differ
from those of the three other galaxies. This is a foreground
group of at least two irregular interacting galaxies situated
at a redshift of $z \sim 0.7$. SEDs of all studied galaxies
except for Id3 are shown in Figure~\ref{sed}b.
%
%
\begin{table*}
 \centering
 \begin{minipage}{150mm}
\caption{parameters of the Id1--5 galaxies determined using \textit{Le Phare}.
Column~1 gives the galaxy ID; column~2 -- the photometric redshift with 95\%
confidence level errors; column~3 -- the corresponding probability; column~4
-- corresponding $\chi^2$ per degree of freedom; column~5 -- the most likely
galaxy SED type (Im -- irregular and interacting galaxies; B -- starburst
galaxies); column~6 gives the best fitted reddening law (LMC -- the law
for Large Magellanic Cloud \citealt{lmc}, SMC -- the law for Small Magellanic
Cloud \citealt{smc}, sb -- the law for starburst galaxies \citealt{calzetti}),
column~7 -- the age of the dominant stellar population of the galaxy in
Gyr; column~8 -- the internal extinction in the galaxy; column~9 -- the
R-band absolute magnitude of the galaxy, and column~10 gives the size of the galaxy in kpc.}
\begin{tabular}{c|c|c|c|c|c|c|c|c|c}
\hline
Id      & $z \pm \Delta z$            & Probability, \% & $\chi^2/DOF$ & Type & Ext.law & Age, & $A_V$,  & M$_R$, & d, \\
        &                             &                 &              & of SED &       & Gyr  & mag     & mag    & kpc$^a$ \\
\hline
1       & $0.73^{+0.10}_{-0.08}$          & $>99.99$ & 46/13 & Im      & sb & 0.50       & 0.28       & $-22.26$          & $49.7 \pm 1.4$ \\
2       & $0.70^{+0.11}_{-0.08}$          & $>99.99$ & 54/13 & Im      & sb & 0.60       & 0.35       & $-21.23$          & $46.3 \pm 2.6$ \\
{\bf 3} & $\mathbf{2.77^{+0.15}_{-0.20}}$ & {\bf 96.32} & 4.0/9 & {\bf B} & LMC & {\bf 0.06}  & {\bf 0.31} & $\mathbf{-22.79}$ & $\mathbf{<7.2 }$ \\
4       & $2.76^{+0.14}_{-0.12}$          & 99.98 & 4.8/9 & B          & sb & 0.02       & 0.49       & $-21.60$          & $<7.2   $\\
5       & $2.84^{+0.16}_{-0.19}$          & 98.23 & 5.0/9 & B          & LMC & 0.05      & 0.03       & $-22.30$          & $<7.2   $\\
\hline
\end{tabular}
\label{z}
\end{minipage}
\end{table*}
%

\citet{evans} used the spectrum of the GRB X-ray afterglow to
determine the hydrogen column density along the line of sight in
the host galaxy of GRB~051008: $N_H = (3.15 \pm 0.35) \times
10^{21}$ cm$^{-2}$ (the provided redshift of the absorber is
equal to 0).
We used the $N_H$--redshift-limit relation from \citet{grupe} to estimate
the upper limit of the GRB~051008 redshift to be $z < 2.8$.
While the upper limit is consistent with our photo-$z$ estimate
one can note that this method is based on a sample
of unobscured bursts, and may be biased toward lower upper limits.
It is known that for optically dark bursts $N_{H}$ values tend to
be systematically higher (e.g., \citealt{fynbo3,kruhler2012}; see also
Table~\ref{GRBs}).

We estimate the star-formation rate (SFR) in the galaxy from its
ultraviolet luminosity according to \citet{kenni}. Using the best
fitted SED and the best host extinction law one can estimate the
UV luminosity of Id3 to be $L_{UV} \sim 4.5 \times 10^{29}$
erg/s/Hz. For the SFR it gives an estimate of $\sim 60$
M$_{\sun}/$yr. The $Le Phare$ package with the PEGASE2 library also
gives the best fitted values of the host physical parameters:
star-formation rate SFR$_{PEGASE2}\sim 60$ M$_{\sun}/$yr and the stellar
mass of the galaxy $M \sim 1.2\times 10^9$ M$_{\sun}$.

In comparison with LBGs with $L^{\ast}$ luminosity Id3 is slightly under-luminous
(by $\sim 0.5$ magnitudes in the rest-frame $V$-band) and has bluer colours.
The age of the dominant stellar population of 60 Myr is 5 times less than the average
value for LBGs, which is not unusual for LBGs: about 25\% of LBGs have
age $ < 40$ Myr \citep{steidel}. Also the Id3 galaxy is less dusty than an average
LBG and 10 times less massive. Its SFR is less than the average LBG SFR,
but close to it.
%
\section{PROPERTIES OF GRB~051008 and its afterglow}

The X-ray afterglow data from
\textit{Swift}/XRT\,\footnote{{\scriptsize{\tt
http://www.swift.ac.uk/xrt\underline{ }curves/00158855/flux.qdp}}}
provides the 2--10~keV flux for the time interval from 3070 to
445845~seconds after the burst onset. Eleven hours after the burst
the X-ray flux was equal to $1.083 \times 10^{-12}$
erg/cm$^{2}$/s. The 3-keV spectral density of the X-ray flux at 11
hours after the trigger, converted in accordance
with~\cite{Jacob}, is $F_X ($3 keV, $11^h) = 0.018 \times
10^{-6}$ Jy. Assuming that the break at $\sim 0.41$ days
(see Section 2.4) is the achromatic jet-break we extrapolated the
optical upper limit of 23\fm3 toward 11 hours after the trigger using
the XRT light curve. We estimate the optical flux to be $F_O (R, 11^h)
< 0.068 \times 10^{-6}$ Jy. From this we derive $\beta_{OX}(11^h)
\leq 0.02$, which allows us to classify this event as a dark burst.
Since the spectral slope in X-ray regime is $\beta_{X} = 1.1$ (see
the end of this Section), this burst is also dark according to the
\citet{vdhorst2} criterion.

We calculated the limit of the isotropic-equivalent radiated energy,
using the redshift value, as $E_{iso} =
(1.13 \pm 0.20) \times 10^{54}$~erg (in the energy range 20 keV -- 10 MeV),
and it is a lower estimate
based on Konus-\textit{WIND} measurements (see Section 2.2).
Performing $k$-correction \citep[e.g.,][]{k-corr}
$E_{iso,\,corrected} = (1.15 \pm 0.20) \times 10^{54}$~erg.
The collimated energy can be estimated given $E_{iso}$ and
the opening angle of the jet. To calculate the jet opening angle
of GRB~051008, we use the corresponding formula from~\citet{sari}
and \citet{liang2}, which relates this parameter to the
isotropic-equivalent radiated energy $E_{iso}$ and the break time
$t_{b} = 0.41^d$ of the afterglow light curve (see Section 2.4):
\begin{equation}
\theta_{j} \sim 0.161 \left(\frac{t_{b}}{1+z}\right)^{3/8}\left(\frac{E_{K,iso,52}}{n}\right)^{-1/8}.
\end{equation}
%
\begin{figure}
\includegraphics[width=84mm]{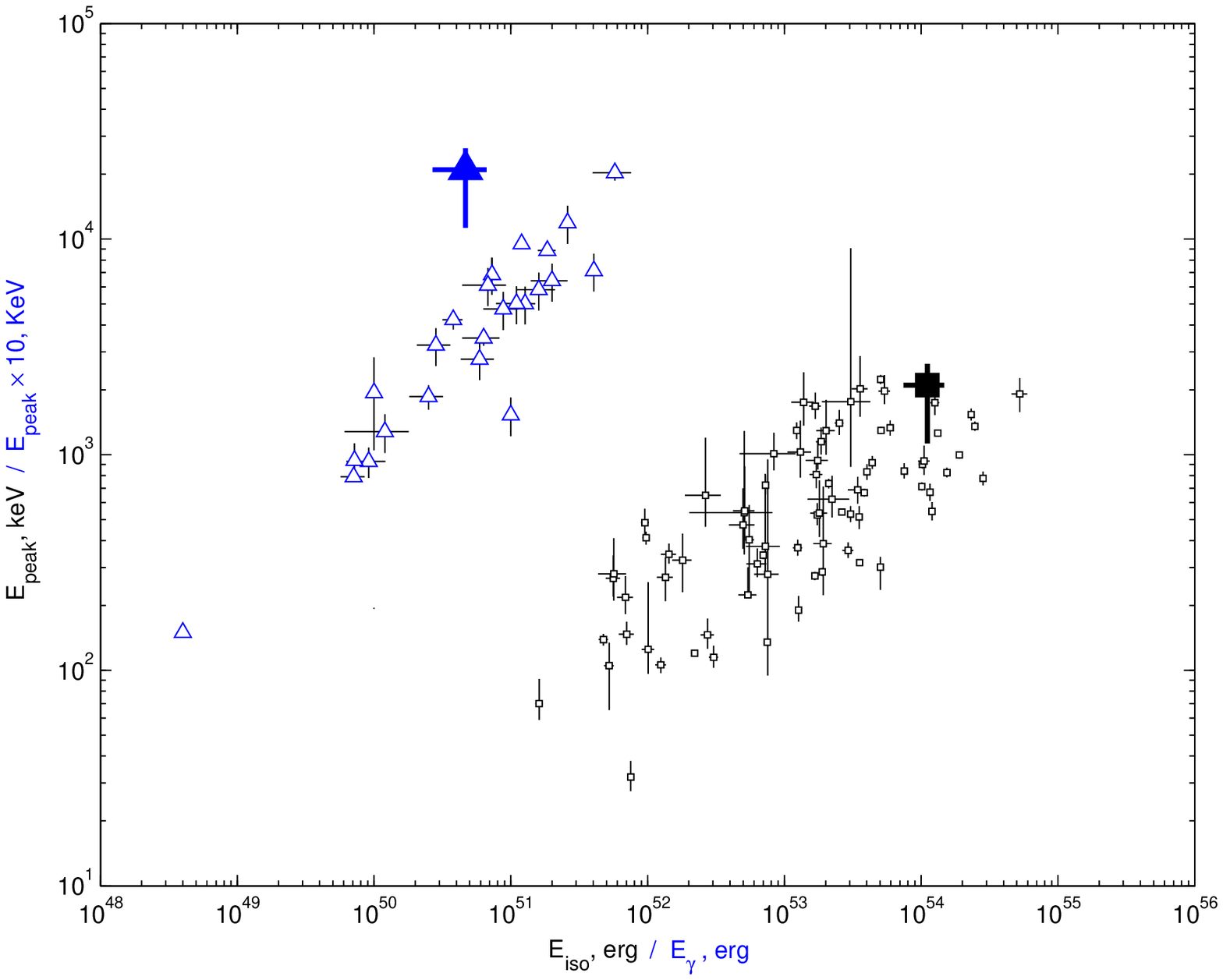}
\caption{The Amati~\citep[squares,][]{amati} and the
Ghirlanda~\citep[triangles,][]{ghirlanda} diagrams. The Amati
diagram displays the dependence of the energy corresponding to the
maximum in the spectrum, $E_p (1+z)$, on the isotropic equivalent
energy, $E_{iso}$. The Ghirlanda diagram displays the dependence
of the energy $E_p (1+z)$ on the collimated energy , $E_{\gamma}$
\citep[adapted from][]{ghirlanda}.
Filled symbols indicate the data points corresponding to
GRB~051008, which can be seen to be in agreement with the Amati
relation, but is a moderately significant outlier of the Ghirlanda
relation assuming $z=2.77$. The Ghirlanda diagram is shifted up
along the Y axis by a factor of 10 for clarity.}
\label{amati-ghirlanda}
\end{figure}

Here $n$ is the constant ISM density, and $E_{K,iso,52}$ is the
total isotropic kinetic energy of the outburst in the units of
$10^{52}$~erg. The latter value is related to the
isotropic-equivalent energy via a formula $E_{iso} = \xi
E_{K,iso}$, where $\xi$ is the coefficient of conversion of the
kinetic energy of the outburst into radiation energy, which
depends on the ambient medium of the GRB source. We assume that
the density of the medium and the conversion coefficient are equal
to $n = 1$ and $\xi = 0.1$ \citep[e.g.,][]{zhang}, respectively.
Given the redshift, the break time of the light curve, and the
equivalent isotropic energy, we estimate the jet opening angle as
$\theta_{j,ISM} = 1^{\circ} .7 \pm 0^{\circ} .2$. Taking these
values into account, the collimation-corrected gamma-ray energy can be
estimated as $E_{\gamma,ISM} = (4.80 \pm 1.52) \times
10^{50}$~erg.

If the GRB progenitor is situated in a wind environment one can
use equation (5) from \citet{chev-li} which connects the jet
opening angle $\theta$ and the parameters of the wind-like medium:
\begin{equation}
\gamma = 5.9 \left(\frac{1+z}{2}\right)^{1/4}E_{52}^{1/4}A_{\ast}^{-1/4}t_{b,days}^{-1/4}, \theta_{j} = 1/\gamma,
\end{equation}
where $E_{52}$ is the isotropic-equivalent energy in units of
$10^{52}$ erg, $t_{b,days}$ is the jet-break time in days, and
$A_{\ast}$ is a parameter of the medium, related to the density
profile as following: $n = Ar^{-2}$, and $A = A_{\ast} \times
5\times 10^{35}$ cm$^{-1}$. Assuming $A_{\ast} = 1$ we can
estimate the value of the jet opening angle $\theta_{j,wind} =
2^{\circ} .0 \pm 0^{\circ} .2$ and corresponding collimated gamma
radiation energy $E_{\gamma,wind} = (7.20 \pm 1.54) \times
10^{50}$~erg.

Using the value of the peak energy in the burst spectrum, we can
plot the parameters of GRB~051008 in the Amati
diagram~\citep{amati}. Data of 95 GRBs used for the Amati diagram
(Fig. \ref{amati-ghirlanda}) were taken from Konus-\textit{WIND}
observations. Additionally, using $\theta_j$ one can put
GRB~051008 in the Ghirlanda diagram~\citep{ghirlanda}, an
empirical diagram showing the correlation between the peak energy
in the spectrum and the emitted $\gamma$-ray energy, i.e.,
$E_{iso}$ corrected for the jet collimation (see
Fig.~\ref{amati-ghirlanda}). While the GRB fits well into the Amati
correlation, it is evident that in case of an ISM environment, it does
not agree with the Ghirlanda correlation. For the possible
explanation see Section 5.1.

We extracted the X-ray spectrum from the raw
\textit{Swift}/XRT data using the XSPEC V12 software (a part of the
HEAsoft software package\footnote{\scriptsize{\tt
http://heasarc.gsfc.nasa.gov/docs/software/lheasoft/}}) for two
different epochs: (a) $t-t_0 =$ 3072 -- 5393 seconds ($\sim 50$
minutes after the trigger) and (b) $t-T_0 =$ 14806 -- 17031
seconds (i.e. $\sim 4$ hours after the trigger). The data were
calibrated using the HEASARC CALibration Data Base
(CALDB\footnote{{\scriptsize {\tt
http://heasarc.gsfc.nasa.gov/FTP/caldb}}}) and fitted with single
power-law with two absorption models
(\textit{phabs*zphabs*powerlaw}). The values of the Galactic
absorption of $N_{H,Gal} = 1.05 \times 10^{20}$ cm$^{-2}$ and the
redshift of $z = 2.77$ were fixed. The fitting provided the value
of the host absorption at the assumed redshift of $N_{H,host}(a) =
(7.7 \pm 1.2) \times 10^{22}$ cm$^{-2}$, $N_{H,host}(b) = (7.9 \pm
1.6) \times 10^{22}$ cm$^{-2}$ and the photon index $\Gamma(a) =
1.8 \pm 0.1$, $\Gamma(b) = 2.1 \pm 0.2$. The value of $N_{H,host}$
is one of the highest known, even for dark bursts (see
Table~\ref{GRBs}). Using this data we can estimate the extinction
along  the line of sight to the GRB source. We construct the
optical SED of the afterglow extrapolating the XRT spectral slope
to the optical bandwidth with a cooling break in the soft end of the X-ray
spectrum (the most conservative estimation). Next we apply the LMC
extinction law in the host galaxy to the model of the SED (see
Section 3). Then we normalize the model of the optical SED to the
upper limit obtained in the same epoch using the early Shajn observation. The
resulting value of the total extinction $A_V^{total}$ in the rest
frame is 3 magnitudes (which effectively is a lower limit for the
total extinction). The intrinsic extinction in the host galaxy is
$A_V^{host} = 0\fm31$. This value may vary significantly
depending on the specific sightline through the galaxy even if we
assume a homogeneous distribution of the absorbing interstellar
medium. For a conservative estimation we
assume $A_V^{host} \sim 1^m$. Thus the source suffers from
additional extinction along the line of sight $A_V^{LOS} > 2^m$.
This indicates that there is an additional absorber in the line of
site.
%
\section{DISCUSSION}
\subsection{Properties of GRB~051008}

\begin{table}
  \caption{Summary of the properties of the host.}
  \begin{tabular}{ll}
\hline
   RA & $13^h31^m29\fs510$ \\
   Dec. & $+42\degr05\arcmin53\farcs67$ \\
   $z_{phot}$ & 2.77 (+0.15,-0.20) \\
    & (95\% confidence level) \\
\hline
   SED type & starburst \\
   $A_{V,bulk}$ & $\sim 0\fm31$ \\
   $M_B$ & $-21\fm6 \pm 0\fm3$ \\
   $M_R$ & $-22\fm8 \pm 0\fm1$ \\
   $M_{K^{\prime}}$ & $-23\fm9 \pm 0\fm3$ \\
   Age & 60 Myr \\
   Mass & $1.2 \times 10^9 M_{\sun}$ \\
   SFR & $\sim 60 M_{\sun}$/y \\
   \hline
   \end{tabular}
  \label{hostprop}
\end{table}

\begin{table}
  \caption{Summary of the GRB and afterglow properties.}
  \begin{tabular}{lll}
   \hline
   $N_{H,z}$ & $(7.9 \pm 1.6) \times 10^{22}$ cm$^{-2}$ & \\
   $E_{iso}$ & $(1.15 \pm 0.20) \times 10^{54}$~erg & \\
   $t_{b}$ & $0.41 \pm 0.14$~days & \\
   $\alpha_1$ & $1.30 \pm 0.18$ & \\
   $\alpha_2$ & $3.18 \pm 0.20$ & \\
   $\beta_{OX}$ (11 hours) & $< 0.02$ & \\
   $\beta_{X}$ (4 hours) & $1.1 \pm 0.2$ & \\
   \hline
    & ISM & wind \\
   $\theta_j$ & $1^{\circ}.7 \pm 0^{\circ}.2$ & $2^{\circ}.0 \pm 0^{\circ}.2$\\
   $E_{\gamma}$ & $(4.80 \pm 1.52) \times 10^{50}$~erg & $(7.20 \pm 1.54) \times 10^{50}$~erg.\\
   \hline
  \end{tabular}
 \label{sourceprop}
\end{table}

We would like to emphasize several properties of GRB~051008. The
absence of statistically significant spectral lag places the burst
in the top 17\% of the Konus-\textit{WIND} burst sample and into
the $\sim 10\%$ of long bursts without spectral lag in the
BATSE~\citep{hakkila} and \textit{Swift}~\citep{ukwatta} samples.
The hardness of the burst, especially the high $E_{peak} = 770$ (CPL)
keV of the main peak of the GRB, which corresponds to $\approx 2.9$ MeV
in the rest frame and places the burst in the 7\% of hardest
time-resolved spectra of the BATSE sample \citep{kaneko}.

GRB~051008 is clearly a long duration burst ($T_{90} = 214$ s
as seen by Konus-\textit{WIND} and $T_{90} = 535$ s as seen by SPI-ACS).
The burst consists of a hard main peak and a somewhat softer tail
detectable up to $\sim$ 800 s. As we have seen from the joint BAT,
SPI-ACS and XRT analysis  the light curve can be described by a
single power-law after the main peak of the burst up to the
jet-break episode. The burst belongs to the less-frequent class of
plateau-less X-ray afterglows \citep{liang3}.

From the modelling of the SED of the host galaxy  one can
conclude that GRB~051008 is optically dark neither because of high
redshift ($z=2.77$), nor due to the global absorption in the host galaxy
$A_V = 0\fm31$. The lack of an OA is likely caused by
absorption in the circumburst medium or by a dense cloud along the line
of sight.

The burst is an outlier of the Ghirlanda relation assuming a value
for the ISM density, $n=1$. An option to make GRB 051008 agree
with the Ghirlanda diagram is that the density of the ISM could be
drastically higher, $n = 10^4-10^6$ instead of $n=1$ used in our
previous estimates. In case of a wind profile, we can estimate the
necessary normalization in the density profile $A$ of $n =
Ar^{-2}$ needed to fulfil the Ghirlanda relation. Let us assume
$E_{\gamma} = 8.2 \times 10^{51}$~erg which would be needed to put
GRB~051008 on the Ghirlanda relation. This yields the value of the
jet opening angle to be $\theta \approx 7\degr$. Using equation
(5) from \citet{chev-li} we obtain the value of $A_{\ast} = 140$
cm$^{-1}$, and $A = A_{\ast} \times 5\times 10^{35}$ cm$^{-1} =
4.2 \times 10^{37}$ cm$^{-1}$. The normalization is $\sim 100$
times greater than the typical wind parameters of Wolf-Rayet stars
\citep[e.g.,][]{castro2010}. This fact supports the hypothesis
that the burst progenitor is embedded in a dense circumburst
medium with a density of $n = 10^4-10^6$ cm$^{-3}$ rather than the
absorber being along the line of sight.

If we suggest a high density of a circumburst medium, one might expect
an appearance of a bright radio counterpart. E.g., for GRB 111215A \citep{zaud}
the maximum  of the radio afterglow light curve at 5.8 GHz was observed
approximately 15 days after the burst at the level of about 1 mJy.
Unfortunately, there were no long-term radio observations of GRB 051008
and deep upper limits (see Section 2.6) are not in contradiction with
a possible radio afterglow evolution similar to that obtained for
GRB~111215A by \citet{zaud}.

Finally, the fact that the redshifts of two further galaxies (Ids 4, 5,
Table~\ref{z}) are close to $z_{Id3}=2.77_{-0.20}^{+0.15}$ may
support the hypothesis that the host galaxy could be
gravitationally paired with at least one other galaxy at a
projected distance of about 50 kpc. The distance between the
galaxies Id3 and Id4 is $46 \pm 10$ kpc, if we adopt a common
redshift of $z=2.77$.
%
\subsection{Comparison with other dark bursts}

\begin{table*}
 \centering
 \begin{minipage}{180mm}
  \caption{A comparison of the parameters of host galaxies of dark GRBs. Column~1 lists the name of the burst
  associated with the galaxy; Columns~2--4 indicate whether the burst had any X-ray, optical/NIR, or radio afterglow
  (''IR'' means the detection of the OA only in the infrared domain, ``n'' means that no observations were made in the
  wavelength interval considered); Column~5 gives the spectral index ($\,^x$ -- the spectral index $\beta_X - 0.5$
  is adopted according to the \citealt{vdhorst2} darkness criterion); Column~6 lists the redshift (superscript $\,^p$
  indicates that the redshift was determined photometrically); Column~7 gives the $R$ magnitude of the host galaxy;
  Column~8 presents the data on the colour of the galaxy; Column~9 gives internal bulk extinction in the galaxy;
  Column~10 -- the extinction towards the GRB source along the line of sight; Column~11 -- the hydrogen column density;
  Column~12 -- the absolute $R$-band magnitude of the galaxy; Column~13 -- the star-formation rate, and Column~14 --
  references: 1 -- \citet{djor}, 2 -- \citet{bloom2}, 3 -- \citet{goros}, 4 -- \citet{piro}, 5 -- \citet{frail},
  6 -- \citet{Jakob2}, 7 -- \citet{kupcu}, 8 -- \citet{levan}, 9 -- \citet{perley2013}, 10 -- \citet{castro},
  11 -- \citet{perley}, 12 -- \citet{tanvir}, 13 -- \citet{jaun}, 14 -- \citet{elias}, 15 -- \citet{svens},
  16 -- \citet{kruhler2012}, 17 -- \citet{hunt}, 18 -- \citet{hashi}, 19 -- \citet{kruhler}, 20 -- \citet{kruhler080605},
  21 -- \citet{chen2010}, 22 -- \citet{holl}, 23 -- \citet{greiner3}, $\ast$ refers to the present paper.}
  {\tiny
  \begin{tabular}{l|ccc|r|r|r|c|r|r|r|r|r|c}
  \hline
  GRB & X & O & R & $\beta_{OX}$ & $z$    & $R_{host}$ & Color    & $A_{V,host}$ & $A_{V,LOS}$ & $N_H \times 10^{21},$ & M$_R$, $^m$ & SFR, & Reference \\
   & & & & & & mag & mag & mag & mag & cm$^{-2}$ & mag & $M_{\sun}/$yr & \\
  \hline
  970828  & $+$ & $-$ & $+$   & $0.09$      & $0.96$ & $25.2$      & $R-K = 3.7$  & ---   & $>3.8$ & $>6$ & $-19.6$ & $\sim 1.2$ & (1) \\
  990506  & $+$ & $-$ & $+$   & $<0.06$   & $1.31$ & $24.3$        & $(R-K)_{AB} = 2.3$ & $>1$  & --- & --- & $-20.5$ & $12.6$      & (2) \\
  000210  & $+$ & $-$   & $+$ & $0.44$      & $0.85$ & $25.0$      & $R-K = 2.5$  & $\sim 0$ & --- & $5 \pm 1$ & $-22.2$  & $\sim 2.1$ & (3,4,5) \\
   & & & & & & & $V-I \sim 1.7$ & & & & & & \\
  020819  & n & $-$ & $+$   & ---       & $0.41$ & $23.46$     & $R-K = 2.7$   & $1.8-2.6$ & $4-20$ & --- & $-22.3$ & $<40$ & (6,7) \\
  030115  & $+$ & $+$ & $+$ & --- & $2.5-2.7^p$ & $25.58^a$ & $R-K = 5.35$ & $\sim 1.0$ & --- & --- & --- & $4.4-500$ & (8) \\
  050915A & $+$ & $-$ & n   & $0.23$ & $2.53$ & $24.56$     & $R-K = 3.9$ & $1.0$ & $\sim 1.4$ & $0.9^{+0.8}_{-0.7}$ & $-22.1$ & $135.8^{+63.1}_{-48.3}$ & (9) \\
  051008  & $+$ & $-$ & $-$ & $< 0.02$ & $2.77^p$ & $24.06$    & $V-I = 0.6$  & $0.31$   & $> 2$ & $79 \pm 16$ & $-22.8$ & $\sim 60$ & ($\ast$) \\
          &   &     &     &           &         &          & $R-K = 1.4$ &        &     &     &            &  & \\
  051022  & $+$ & $-$ & $+$   & $<-0.11$  & $0.81$ & $21.5$      & $R-K = 3.3$  & $1.0$   & $> 9$ & $34.7^{+4.8}_{-4.7}$ & $-21.8$  & $\sim 50$ & (9,10) \\
   & & & & & & & $V-I \sim 1.1$ & & & & & \\
  060202  & $+$ & $-$ & n   & $< 0.20$ & $0.78$ & $23.79^b$ & $(R-K)_{AB} = 1.1$ & $\sim 1$ & $2.9$ & $4.5^{+0.7}_{-0.6}$ & $-19.6$ & $5.8^{+1.3}_{-1.9}$ & (9) \\
  060210  & $+$ & $+$ & $-$ & $0.37$   & $3.91$ & $24.33^b$  & $(R-I)_{AB} \approx 0.1$   & $0.25$  & $1.2^{+0.2}_{-0.1}$ & $24.6^{+2.9}_{-3.5}$ & $<-20.2$ & $-23.4$ & (11) \\
  060306  & $+$ & $-$ & n   & $< 0.21$ & $1.55$ & $24.49^b$ & $(R-K)_{AB} = 2.5$ & $2.2 \pm 0.1$ & $4.6$ & $3.0^{+0.8}_{-0.7}$ & $-20.8$ & $244^{+129}_{-67}$ & (9) \\
  060319  & $+$ & $+$ & n   & $0.19$ & $1.17$ & $24.20^b$ & $(R-K)_{AB} = 2.4$ & $0.1^{+0.2}_{-0.1}$ & $1.1$ & $3.4^{+0.8}_{-0.7}$ & $-20.3$ & $0.0^{+0.4}_{-0.0}$ & (9) \\
  060719  & $+$ & $+$ & $-$ & $-0.27$ & $1.53$ & $24.62^b$ & $(R-K)_{AB} = 1.8$ & $0.4^{+1.1}_{-0.4}$ & $2.1$ & $3.4^{+1.3}_{-1.0}$ & $-20.6$ & $4.0^{+36.9}_{-4.0}$ & (9) \\
  060814  & $+$ & $+$ & n   & $-0.21$ & $1.92$ & $22.85$ & $(R-K)_{AB} = 1.2$ & $1.2^{+0.1}_{-0.0}$ & $3.1$ & $2.7^{+0.4}_{-0.3}$ & $-23.0$ & $236.7^{+28.4}_{-18.3}$ & (9) \\
  060923A & $+$ & $+$   & $-$ & $0.38$    & $<2.8^p$ & $25.65$      & $R-K \sim 4$ & --- & $2.6$ & ---     & $<-21.2$ & --- & (11,12) \\
           &   &     &    &           &         &          & $V-I \sim 1.5$ &      &  &          &   &         & \\
  061222A & $+$ & $+^{IR}$ & $+$ & $<-0.19$ & $2.09$ & $24.93$ & $(V-I)_{AB} \sim 0.5$ & $< 0.5$ & $> 5.0$ & $44.8^{+5.4}_{-3.0}$ & $-21.2$ & --- & (9,11) \\
          &   &       &   &       &      &     & $(I-K)_{AB} < 1.0$    &       &  &     &     & & \\
  070306  & $+$ & $+^{IR}$ & $-$ & $0.33$ & $1.50$ & $\sim 23$ & $R-K \sim 1.5$ & $\leq 0.45$ & $5.5 \pm 0.6$ & $26.8^{+4.7}_{-4.3}$ & $-22.2$ & $\sim 7.3$ & (13) \\
  070521  & $+$ & $-$   & n & $< -0.10$      & $1.35$ & $26.11$  & $V-I = 0.72$    & ---   & $> 9$ & $54^{+13}_{-11}$ & $-18.8$ & --- & (11) \\
  070802  & $+$ & $+$   & n   & $0.46$      & $2.45$ & $25.03$     & $R-K \sim 3.3$ & $<1.5$ & $<0.3$ & $3.2 \pm 1.8$ & $-21.0$ & --- & (14) \\
  071021  & $+$ & $+^{IR}$   & $+$ & $0.37$ & $2.45$ & $25.22^b$   & $(R-K)_{AB} = 3.1$ & $1.6 \pm 0.3$ & $0.5$ & $0.5^{+0.8}_{-0.5}$ & $-21.3$ & $108^{+98}_{-55}$ & (9) \\
  080207  & $+$ & $-$   & n & $< 0.3$  & $2.08$ & $>25.65$ & $R-K > 5.4$ & $\sim 1.9$ & $\leq3.4$ & $151^{+23}_{-22}$ & $>-19.9$ & $\sim 40$ & (15,16) \\
   & & & & & $\sim 2.2^p$ & $\sim 25.8$ & $R-K \sim 6.3$ & $1-2$ & --- & & & $\sim 119$ & (17) \\
  080325  & $+$ & $+^{IR}$ & n & $0.14-0.33$ & $1.78$ & $25.5$ & $R-K_s = 3.8$ & $0.8$ & $2.7-10$ & $18^{+8}_{-6}$ & $-20.3$ & $10-80$ & (18,9) \\
  080605  & $+$ & $+$   & n   & $0.17^x$ & $1.64$ & $21.6$ & $(R-K)_{AB} = 0.5$ & $0.5\pm0.1$ & $1.2\pm0.1$ & $9.0 \pm 0.9$ & $-22.4$ & $49^{+26}_{-13}$ & (19,20) \\
  080607  & $+$ & $+$   & n   & $0.14-0.37$ & 3.04 & $26.33^b$ & $(R-K)_{AB} = 2.5$ & $1.25$ & $3.3$ & $1.4 \pm 0.4$ & $-20.8$ & $18.3^{+5.0}_{-3.2}$ & (9,21) \\
  081109A  & $+$ & $+$ & $-$ & $<0.44$ & $0.97$ & $22.65$ & $(R-K)_{AB} = 1.6$ & $1.0\pm0.2$ & $3.4^{+0.4}_{-0.3}$ & $9 \pm 2$ & $-21.4$ & $33^{+19}_{-13}$ & (19) \\
  081221  & $+$ & $+$ & $+$ & $0.00$ & $2.26$ & $24.67^b$ & $(R-K)_{AB} = 2.5$ & $1.4\pm0.1$ & $1.4$ & $3.9^{+0.5}_{-0.4}$ & $-21.6$ & $73^{+16}_{-13}$ & (9) \\
  090404  & $+$ & $-$ & $+$ & $<0.20$ & $3.0^p$ & $26.07^b$ & $(R-K)_{AB} = 3.1$ & $1.3^{+2.2}_{-1.2}$ & $1.3$ & $5.1^{+1.0}_{-0.9}$ & $-21.0$ & $40^{+178}_{-22}$ & (9) \\
  090407  & $+$ & $-$ & n & $<0.14$ & $1.45$ & $25.78^b$ & $(R-K)_{AB} = 3.3$ & $1.8^{+0.1}_{-1.2}$ & $1.6$ & $2.3\pm0.4$ & $-19.3$ & $16.6^{+1.8}_{-15.2}$ & (9) \\
  090417B & $+$ & $-$ & $-$ & $<0.28$ & $0.34$ & $21.34$ & $R-K = 2.8$ & $0.9\pm0.1$ & $2.7$ & $8.3^{+1.4}_{-1.2}$ & $-20.0$ & $0.5\pm0.3$ & (9,22) \\
  090709A & $+$ & $-$ & $-$ & $-0.56$ & $1.8^p$ & $26.39^b$ & $(R-K)_{AB} = 2.7$ & $1.4^{+0.1}_{-0.2}$ & $3.0$ & $1.7\pm0.3$ & $-19.3$ & $8.3^{+4.4}_{-3.6}$ & (9) \\
  090926B & $+$ & $+$   & n   & $0.23^x$ & $1.24$ & $22.94$ & $(R-K)_{AB} = 1.5$ & $1.4\pm0.3$ & $1.4^{+1.1}_{-0.6}$ & $13.9^{+1.6}_{-1.5}$ & $-21.8$ & $80^{+110}_{-50}$ & (19) \\
  100621A  & $+$ & $+$   & n   & $0.39$ & $0.542$ & $21.53$ & $(R-K)_{AB} = 0.3$ & $0.6\pm0.1$ & $3.8\pm0.2$ & $18.0^{+1.2}_{-1.1}$ & $-20.9$ & $13^{+6}_{-5}$ & (19,23) \\
  \hline
  \end{tabular}
  }
 \label{GRBs}
 \newline $^a${\tiny{in $F606W$ filter.}}
 \newline $^b${\tiny{synthetic AB magnitude.}}
 \end{minipage}
\end{table*}

Table~\ref{GRBs} lists the parameters of dark bursts for which
host galaxies have been found. The table partially
overlaps with the dark burst sample presented by
\citet{perley2013}. Our table lists the following parameters of
the galaxies: redshift, apparent and absolute $R$-band magnitudes,
colour information, intrinsic host absorption, absorption along the
line of sight to the GRB source, hydrogen column density, and
star-formation rate. It comes as no surprise that the GRBs
presented in Table~\ref{GRBs} exhibit high line-of-sight
extinction, since the assumption employed in selecting this sample
(i.e., a redshift that does not point to the influence of Lyman
absorption, and the application of a single synchrotron spectrum
in modelling the broad-band SED) leave extinction as the strongest
choice to explain the optical/NIR darkness
\citep{kruhler2012,perley2013}. We also find that many dark-burst
host galaxies exhibit a high SFR, with the median SFR for $z>1$
hosts in our list being $\approx40$ $M_{\sun}/$yr. Indeed,
GRB~051008 is one of the most distant bursts for which a host galaxy
has been investigated in detail.

\citet{greiner} analysed the extinction along the line of sight
through (undetected) host galaxies of 33 GRBs, nine of which are
dark bursts. The values of $A_V$ were obtained directly from
observed SEDs of the afterglows (X-ray and optical). They found
that a significant fraction of the bursts (25\%) have $A_V \sim
0\fm5$ and another 10\% have $A_V > 1^m$. GRB~051008 with its
$A_{V,LOS} > 2^m$ is located in the 10\% of the bursts with high
$A_V$ \citep{zafar}.

\subsection{Conclusions}

In this paper we gathered almost all available observations of the
dark GRB~051008, including $\gamma$-ray and X-ray data, optical
imaging and spectroscopy, NIR, and radio data. GRB~051008 is
clearly an optically dark burst.\, No OA was detected for GRB~051008
to a limit of $23\fm3$ within half an hour after the burst
trigger, and the burst is dark in accordance with the darkness
criteria of both \citet{Jacob} and \citet{vdhorst2}. We detected
and studied the host galaxy of GRB~051008. It is a rather typical
Lyman-break galaxy located at the redshift of $z_{phot} = 2.77_{-0.20}^{+0.15}$.
The host galaxy has a SFR of $\sim 60 M_{\sun}/$ yr typical for LBGs. The host is
slightly younger and less dusty than LBGs with $L^{\ast}$ luminosity, 
and it is almost
10 times less massive. It is one of the few cases where a GRB
host has been found to be a classical Lyman-break galaxy.

The redshift and the detection of GRB~051008 with
Konus-\textit{WIND} allowed us to calculate the
isotropic-equivalent energy of the burst to be $E_{iso} = (1.15
\pm 0.20) \times 10^{54}$~erg. This is a highly energetic long
burst without detectable spectral lag and with a hard spectrum
and rather early jet-break of $\sim 0.41$ days determined from
X-ray observations. The study of the host galaxy infers the presence
of additional line of sight extinction towards the burst source.
This extinction can not be explained by the bulk extinction in
the host galaxy, which is only moderate. Most probably it is produced
by a dense circumburst medium.

We furthermore found the
photometric redshift of nearby galaxies, and among them the two
galaxies have a redshift $z\sim2.8$, coinciding with the
redshift of the host within statistical uncertainties. Therefore,
the host may be situated in a gravitationally bound system. The
redshift of all three galaxies is determined due to the presence
of a clear and strong Lyman break feature.
%
\section*{acknowledgments}

We are grateful to S.N.~Dodonov, T.A.~Fatkhullin, M.V.~Barkov,
O.V.~Egorov, M.~Salvato, D.I.~Karasyov, and  V.A.~Kolesnikov for
useful discussions. A.A.V., A.S.P. and P.Yu.M. were supported by
the program "Origin, structure, and evolution of objects in the
Universe" funded by the Russian Academy of Sciences and RFBR
grants 12-02-01336-a, 13-01-92204-Mong, 14-02-10015-K.

J.G. and A.J.C-T.  were supported by the Spanish programs
ESP2005-07714-C03-03,  AYA2007-63677,  AYA2008-03467/ESP, and
AYA2009-14000-C03-01. The data presented in this paper have been
acquired using the ALFOSC camera, which is owned by the Instituto
de Astrof\'{i}sica de Andaluc\'{i}a (IAA) and operated at the
Nordic Optical Telescope under the agreement between IAA and
NBIfAFG of the Astronomical Observatory of Copenhagen.
The Konus-\textit{WIND} experiment is partially supported by a
Russian Space Agency contract, RFBR grants 12-02-00032a and 13-02-12017 ofi-m.

D. A. K. and S. K. acknowledge financial support by DFG grants Kl
766/13-2 and Kl 766/16-1. D. A. K. acknowledges financial support by
MPE and TLS.

Support for D.A.P. is provided by NASA through Hubble Fellowship
grant HST-HF-51296.01-A awarded by the Space Telescope Science
Institute, which is operated for NASA by the Association of Universities
for Research in Astronomy, Inc., under contract NAS 5-26555.

\bibliographystyle{mn2e}

\label{lastpage}
\end{document}